\documentclass[letterpaper,prb,twocolumn,showpacs]{revtex4}
\usepackage{graphicx,bm}
\begin{document}

\title{Identifying the pairing symmetry in sodium cobalt oxide by Andreev edge states}
\author{Wen-Min Huang}
\author{Hsiu-Hau Lin}
\affiliation{Department of Physics, National Tsing-Hua University, Hsinchu 300, Taiwan}
\affiliation{Physics Division, National Center for Theoretical Sciences, Hsinchu 300, Taiwan}

\date{\today}

\begin{abstract}

We study the Andreev edge states with different pairing symmetries and boundary topologies on semi-infinite triangular lattice of Na$_x$CoO$_2$$\cdot y$H$_2$O. A general mapping from the two dimensional lattice to the one dimensional tight-binding model is developed. It is shown that the phase diagram of the Andreev edge states depends on the pairing symmetry and also the boundary topology. Surprisingly, the structure of the phase diagram crucially relies on the nodal points on the Fermi surface and can be explained by an elegant gauge argument. We compute the momentum-resolved local density of states near the edge and predict the hot spots which are measurable in Fourier-transformed scanning tunneling spectroscopy.

\end{abstract}
\maketitle

\section{introduction}

The recent discovery of superconductivity in sodium cobalt oxide compound intercalated water molecules,  Na$_x$CoO$_2$$\cdot y$H$_2$O,\cite{Takada03} trigged intense attentions and stimulated lots of discussions\cite{Ogata07}. The superconductivity induced in the planer structure of CoO$_2$ is similar with that in the CuO$_2$ plane of cuprates\cite{Chou04,Ihara04}. However, the underlaying triangular lattice of the Co atoms is fundamentally different from the square lattice of the Cu atoms in cuprates because the antiferromagnetic interactions on the triangular lattice are frustrated. The carrier density in the sodium cobalt oxide can be tuned by the Na concentration. By changing the sodium doping, a rich phase diagram appears and the superconductivity occurs\cite{Schaak03,Foo04} in the doping regime $1/4<x<1/3$. Furthermore, the study in Co-NMR and Co-NQR found that the spin-lattice relaxation rate at the critical temperature ($T_c$) shows no coherent peak and follows a power below $T_c$, hiniting an unconventional superconducting phase\cite{Ishida03,Fujimoto04,Zheng061}. The node of the superconducting gap is confirmed by the specific-heat measurements\cite{Yang05} and also by the muon spin relaxation experiments\cite{Kanigel04}. 

However, the symmetry of the Cooper pairs remains unknown at present. In order to identify the pairing symmetry, the measurement of spin susceptibility in the superconducting state through the Knight shift is helpful\cite{Higemoto04,Kobayashi05,Ihara06}. The measurements of the powder samples show that the Knight shifts along the $c$-axis do not decrease below $T_c$, raising the possibility of spin-triplet superconducting state\cite{Kobayashi05,Ihara06,Kato06}. On the other hand, recent measurements on the single-crystal samples\cite{Zheng062} show that the Knight shift decreases below $T_c$ along the $a$- and $c$- axes, which suggests for the spin-singlet pairing instead. From the study of the normal-state Fermi surface topology by the angle-resolved photoemission spectroscopy\cite{Shimojima06} and the Mn doping effects\cite{Chen07}, it also seems to support the singlet superconducting state. Thus, the pairing symmetry of superconductivity in Na$_x$CoO$_2$$\cdot y$H$_2$O compounds remains controversial at the point of writing.

There are also theoretical efforts to pin down the pairing symmetry of the gap function in Na$_x$CoO$_2$\cite{Mazin05}. The underlaying triangular lattice is proposed to host the resonating-valence-bond (RVB) state for an unconventional superconductor\cite{Baskaran03}. Base on the RVB picture, theoretical investigations on the $t$-$J$ model\cite{Kumar03,Wang04} favor the $d_{x^2-y^2}+id_{xy}$ symmetry. However, within the third-order perturbative expansions, a stable $f$-wave pairing is found in the Hubbard model\cite{Ikeda03} with repulsive on-site interaction. The same conclusion is reached from the theoretical study on the single-band extended Hubbard model within random phase approximations\cite{Tanaka032}. Furthermore, recent discovery of the Hubbard-Heisenberg model on the half-filled anisotropic triangular lattice show that varying the frustration $t'/t$ changes the spatial anisotropy of the spin correlations and leads to transitions of the pairing symmetries of the superconducting oder parameter\cite{Powell07}. Taking different routes for theoretical investigations, other groups demonstrate the possibility of the $p_x+ip_y$ pairing\cite{Tanaka03,Han041}. In addition, starting from the fluctuation-exchange approximations, the triplet $f$-wave and $p$-wave pairings are favored on the triangular lattice\cite{Mochizuki05}. With the same approximations, solving the linearized {\'E}liashberg equation\cite{Kuroki04} leads to dominant pairing in the spin-triplet $f$-wave sector. Therefore, the pairing symmetry also posts a challenging task for theoretical understanding from the microscopic perspective. 

While it is important to determine the pairing symmetry from microscopic approaches, it is equally crucial to develop phenomenological theories so that one can extract the pairing symmetry from the experimental data\cite{Han042,Li04} such as the Andreev bound states\cite{Braunecker05,Lin05} near the edges of the superconductors. Note that the Andreev edge state\cite{Hu94} in a superconductor is tied up with the pairing symmetry in the bulk. In addition, recent breakthroughs in the Fourier-transformed scanning tunneling spectroscopy (FT-STS) experiments\cite{Hoffman02,McElroy03} allow further insight into the edge states with momentum resolutions. In these experiments, not only the spatial profile of the local density of states (LDOS) can be measured, the peaks of the LDOS in the momentum space can also be determined by appropriate Fourier analysis of the experimental data. In a letter published by one of the authors\cite{Lin05}, a theoretical approach was developed to compute the momentum-resolved LDOS for the Andreev edge state in sodium cobalt oxide with $f$-wave pairing symmetry. The exponential decay away from the boundary can be compared with the experiments directly, while the dependence upon the transverse momentum (along the edge where the system is translational invariant ) can be seem in Fourier space through scattering processes. Here, we elaborate and extend the previous work by considering gap functions of $p$-, $d$- and $f$-pairing at both zigzag and flat edges and predict the position of the sharp peaks that can be observed in FT-STS experiments.

\begin{table}
\begin{tabular}{ c c c } \hline \hline
\hspace{0.4cm}Zigzag edge \hspace{0.4cm}& \hspace{0.4cm}Flat edge\hspace{0.4cm} & \hspace{0.4cm}Pairing symmetry\hspace{0.4cm} \\ \hline
yes & no & $p_x$ or $f$\\ \hline
no & yes & $p_y$ \\ \hline
yes & yes & $d_{xy}$ \\ \hline
no & no & $d_{x^2-y^2}$ or $s$ \\ \hline \hline
\end{tabular}
\caption{Existence of Andreev edge state at zigzag and flat edges and its implication for pairing symmetry.}
\end{table}

We start with the two dimensional (2D) Bogoliubov-de Gennes Hamiltonian and map the semi-infinite triangular lattice to a collection of one-dimensional (1D) chains, labeled by the transverse momentum along the boundary. Due to the hidden structure of these effective 1D models, the AES can be categorized into the positive and negative Witten parity states\cite{SUSY} in supersymmetric (SUSY) algebra. For readers no familiar with the Witten parity and the SUSY algebra, we have included a brief introduction in Appendix A. By computing the Witten parity states constrained by the boundary conditions, the LDOS with specific transverse momentum is obtained. Furthermore, we can predict the hot spots in FT-STS by spotting all momentum differences between sharp peaks in the LDOS. Our results show that the existence of AES sensitively depends on the pairing symmetry and the edge topology and can thus be used as a good indicator of the underlying pairing symmetry. The existence of the AES for different pairing symmetries and edge topologies are summarized in Table. I. Finally, following an elegant gauge argument devised by Oshikawa\cite{Oshikawa00,Refael05}, we also find that the phase diagram for the AES crucially depends on the nodal points on the Fermi surface where the pairing amplitude vanishes.

The rest of paper is organized as the followings. In Sec. II, we introduce the 2D Bogoliubov-de Gennes Hamiltonian for a triangular lattice with the zigzag boundary topology. By transforming the Hamiltonian into Supersymmetric form and use the generalized Bloch state, the LDOS of AES is obtained. In Sec. III, in the same spirit and method, we will compute the LDOS of AES for the flat edge. We will discuss about the gauge argument of phase diagram and draw a conclusion in Sec. IV. 

\section{Bogoliubov-de Gennes Hamiltonian at zigzag edge}

To accommodate different pairing symmetries within one theoretical framework, it is convenient to start from the Bogoliubov-de Gennes (BdG) Hamiltonian\cite{deGennes},
\begin{eqnarray}\label{BdG}
\hspace{-0.4cm}H_{BdG}=&&\hspace{-0.3cm}t\sum_{\langle \textbf{r},
 \textbf{r}'\rangle,\sigma}c^{\dag}_{\sigma}( \textbf{r})c_{\sigma}( \textbf{r}')-\mu\sum_{ \textbf{r},\sigma}c^{\dag}_{\sigma}( \textbf{r})c_{\sigma}( \textbf{r})\nonumber
\\ &&\hspace{-1.3cm}+\sum_{\langle  \textbf{r},
 \textbf{r}'\rangle}\left[\Delta^*( \textbf{r}, \textbf{r}')c_{\uparrow}( \textbf{r})c_{\downarrow}( \textbf{r}')+\Delta( \textbf{r}, \textbf{r}')c^{\dag}_{\downarrow}( \textbf{r}')c^{\dag}_{\uparrow}( \textbf{r})\right],
\end{eqnarray}
where only the nearest-neighbor hopping and pairing are included. Because the particle-hole symmetry is absent in the triangular lattice, the sign of the hopping amplitude $t$ is crucial. Recent experiments\cite{Singh,Valla} suggest that the maximum of the band occurs at the $\Gamma$ point, which implies $t>0$. The pairing amplitudes are either symmetric $\Delta( \textbf{r}, \textbf{r}')=\Delta( \textbf{r}', \textbf{r})$ or antisymmetric $-\Delta( \textbf{r}', \textbf{r})$ depending on the Cooper pairs are spin singlets or triplets.

In this paper,  we will discuss two natural boundary topologies of a triangular lattice -- zigzag and flat edges, as showed in Fig. \ref{PSZf} and Fig. \ref{PSFd} respectively. The conventions for the spatial coordinates and also the pairing symmetries can be found in the figures as well. For instance, the zigzag edge is chosen to lie in the $y$-axis and the flat edge along the $x$-axis in our convention.

We start with the zigzag edge first, by cutting the infinite triangular lattice along the $y$-axis. Note that the semi-infinite lattice is still translational invariant along the boundary and thus can be mapped onto a collection of semi-inifinite 1D chains, carrying definite transverse momentum $k_y$ after partial Fourier transformation. One important subtlety about partial Fourier transformation is the folding of Brillouin zone. The conventional hexagonal shape must be reshaped into appropriate rectangular one so that the summations over $k_x$ and $k_y$ are decoupled\cite{LIN98}. For the zigzag edge, the reconstruct rectangle Brillouin zone is shown in the bottom of Figs. \ref{PSZf}, \ref{PSZdxy} and \ref{PSZpx}. After the partial Fourier transformation, the Hamiltonian for the collection of the effective 1D chains along $x$-direction is
\begin{eqnarray}\label{BdGZ}
H=\sum_{k_{y}}\Phi^{\dag}(k_{y}) \left(
\begin{array}{cc}
\mathcal{H} & \mathcal{P}_{\nu} \\
\mathcal{P}^{\dag}_{\nu} & -\mathcal{H}
\end{array}\right)
\Phi(k_{y}),
\end{eqnarray}
with $\nu=f,d$ and $p$, which are denoted as $f$-, $d$- and $p$-wave pairing respectively. Here we introduce the Nambu basis $\Phi^{\dag}(k_{y})=\left[c^{\dag}_{\downarrow}(x,k_{y})\hspace{0.1cm},\hspace{0.1cm}c_{\uparrow}(x,-k_{y})\right]$ and the semi-infinite matrix for the hopping term of semi-infinite 1D chains, 
\begin{eqnarray}
\mathcal{H}=\left(
\begin{array}{cccccc}
-\mu & t_1& t_2 & 0 & 0 & \cdots \\
t_1 & -\mu & t_1 & t_2 & 0 & \cdots \\
t_2 & t_1 & -\mu & t_1 & t_2 & \cdots \\
0& t_2 & t_1& -\mu & t_1& \cdots \\
\cdot & \cdot & \cdot & \cdot & \cdot & \cdots \\
\cdot & \cdot & \cdot & \cdot & \cdot & \cdots
\end{array}
\right),
\end{eqnarray}
with the effective hopping amplitude $t_1=2t\cos(\sqrt{3}k_y/2)$ and $t_2=t$. The momentum dependence of the matrix elements is a consequence of the partial Fourier transformation. The pairing potential $\mathcal{P}_{\nu}$ with different symmetries will be studied in details in the following subsections.

\subsection{$f$-wave paring}

\begin{figure}
\begin{center}
\includegraphics[width=6.2cm]{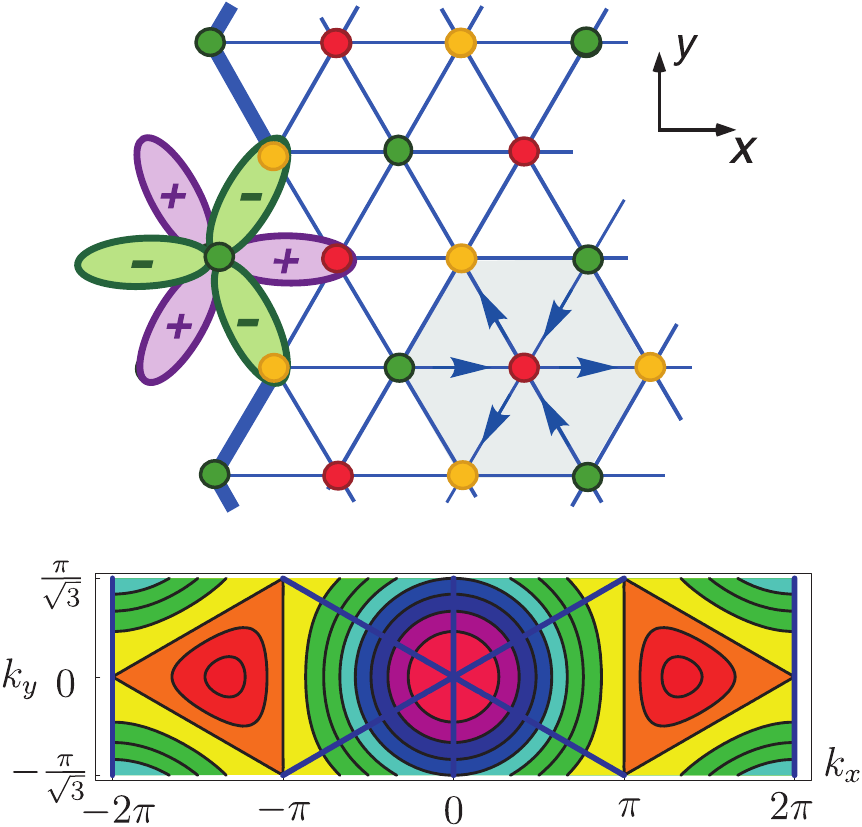}
\caption{(Color Online) Gap function with the $f$-wave symmetry at the zigzag edge of a triangular lattice. The sign convention of the pairing potential is shown in the shaded hexagon. The bottom figure represents the Fermi surface in the reshaped Brillouin zone for the zigzag edge. The nodal lines of the $f$-wave gap function are shown in blue lines and the nodal points are the intersections of the Fermi surface contour and the nodal lines. }\label{PSZf}
\end{center}
\end{figure}

We start with the AES of $f$-wave pairing symmetry at zigzag edge. The $f$-wave symmetry carries angular momentum $l=3$ and thus corresponds to spin-triplet pairing required by Fermi statistics. It implies that the pairing potential is antisymmetric, $\Delta(\textbf{r}, \textbf{r}')=-\Delta( \textbf{r}', \textbf{r})$. Taking the tight-binding approximation, the pairing potential is rather simple $\Delta( \textbf{r}, \textbf{r}')=\Delta(\theta)=\Delta\cos(3\theta)$ with the relative angle $\theta=2n\pi/6$ where $n$ is an integer. The sign convention for different bond orientations is fixed in Fig. \ref{PSZf}. We can solve for the nodal lines by setting the gap function to zero, $\left[\cos(k_x/2)-\cos\left(\sqrt{3}k_y/2\right)\right]\sin(k_x/2)=0$. These nodal line are drawn in the reshaped Brillouin zone in Fig. \ref{PSZf}. At different fillings (chemical potentials), the nodal points are the intersections of the Fermi surface contour and the nodal lines. These nodal points turn out to be the key for determining the structure of the phase diagrams for the AES. The presence of the open boundary complicates the story and we need to write down the pairing potential in the coordinate space. After some algebra, the semi-infinite matrix $\mathcal{P}_{f}$ of Eq. (\ref{BdGZ}) takes the form,
\begin{eqnarray}
\mathcal{P}_{f}=\left(
\begin{array}{cccccc}
0 & -\Delta_1 & \Delta & 0 & 0 & \cdots \\
\Delta_1& 0 & -\Delta_1 & \Delta & 0 & \cdots \\
-\Delta & \Delta_1 & 0 & -\Delta_1 & \Delta & \cdots \\
0& -\Delta & \Delta_1 & 0 & -\Delta_1 & \cdots \\
\cdot & \cdot & \cdot & \cdot & \cdot & \cdots \\
\cdot & \cdot & \cdot & \cdot & \cdot & \cdots
\end{array}
\right),
\end{eqnarray}
with $\Delta_1=2\Delta\cos\left(\sqrt{3}k_y/2\right)$. A simple unitary transformation brings the Hamiltonian into SUSY form described in Appendix A. The effective Hamiltonian\cite{SUSY,Huang04} in canonical SUSY notation is
\begin{eqnarray}\label{DiracZ}
H=\sum_{k_{y}}\Psi_{\nu}^{\dag}(k_{y}) \left(
\begin{array}{cc}
0 & \mathcal{A}_{\nu} \\
\mathcal{A}_{\nu}^{\dag} & 0
\end{array}\right)
\Psi_{\nu}(k_{y}),
\end{eqnarray} 
where the matrix $\mathcal{A}_{\nu}$ takes the general form 
\begin{eqnarray}
\mathcal{A}_{\nu}=\left(
\begin{array}{cccccc}
-\mu & T_{1}^{\nu} & T_{2}^{\nu} & 0 & 0 & \cdots \\
T_{\bar{1}}^{\nu} & -\mu & T_{1}^{\nu} & T_{2}^{\nu} & 0 & \cdots \\
T_{\bar{2}}^{\nu} & T_{\bar{1}}^{\nu} & -\mu & T_{1}^{\nu} & T_{\bar{2}}^{\nu} & \cdots \\
0& T_{\bar{2}}^{\nu}& T_{\bar{1}}^{\nu} & -\mu & T_{1}^{\nu} & \cdots \\
\cdot & \cdot & \cdot & \cdot & \cdot & \cdots \\
\cdot & \cdot & \cdot & \cdot & \cdot & \cdots
\end{array}
\right).
\end{eqnarray}
Although we concentrate on the $f$-wave symmetry in this section, the derivations of the matrix elements of $\mathcal{A}_{\nu}$ are completely general and work for different  pairing symmetries. 

For current case, $\mathcal{P}^{\dag}_{f}=-\mathcal{P}_{f}$ for $f$-wave pairing, the new basis for the SUSY form is 
\begin{eqnarray}\label{fbasis}
\Psi_{f}(k_{y})= \left[
\begin{array}{c}
c_{\downarrow}(x,k_{y})- c_{\uparrow}^{\dag}(x,-k_{y})\\
c_{\downarrow}(x,k_{y})+ c_{\uparrow}^{\dag}(x,-k_{y}) 
\end{array}\right],
\end{eqnarray}
and the matrix elements of $\mathcal{A}_{f}$ are
\begin{eqnarray}
\nonumber T_{1,\bar{1}}^f&=&2(t\mp\Delta)\cos\left(\frac{\sqrt{3}}{2}k_y\right),\\
T_{2,\bar{2}}^f&=&t\pm\Delta.
\end{eqnarray}
It will become clear later that $T_{1}^f\neq T_{\bar{1}}^f$ and $T_{2}^f\neq T_{\bar{2}}^f$ are the crucial for the existence of the edge states.

For the Hamiltonian in Eq.~(\ref{DiracZ}), the zero-energy states are ``nodal", i.e. half of the components in the spinor vanish, and can be classified by the so-called SUSY parity (see Appendix A),
\begin{eqnarray}\label{nodalsol1}
|\Psi_{-}\rangle= \left(
\begin{array}{c}
0 \\ \psi_{-}(x)
\end{array} \right),
\hspace{0.5cm} |\Psi_{+}\rangle=\left(
\begin{array}{c}
\psi_ {+}(x) \\ 0
\end{array} \right).
\end{eqnarray}
It is straightforward to show that the Harper equations become decoupled for the zero-energy states and simplify a bit. The solution with positive Witten parity $\psi_{+}(x)$ is annihilated by $\mathcal{A}_{\nu}^{\dag}$, i.e. it belongs to the null space of the operator. Similarly, the solution with negative Witten parity $\psi_{-}(x)$ spans the null space of the operator $\mathcal{A}_\nu$. It is worth emphasizing that bring the Hamiltonian into the SUSY form simplifies the algebra and allows analytic calculations for AES as derived here. 

To include the open boundary condition, the edge state can be constructed by the generalized Bloch theorem\cite{Lin05}. Taking states with negative Witten parity as a working example, one can construct an edge state from appropriate linear combinations of the zero-energy modes, $\psi_{-}(x)=\sum_{\gamma} a_{\gamma}(z_{\gamma})^x$. Since the zero energy modes satisfy $\mathcal{A}_{\nu}\psi_{-}(x)=0$, $z$ is a solution of the following characteristic equation,
\begin{equation}\label{zsol1}
T_{\bar{2}}^{\nu}\frac{1}{z^2}+T_{\bar{1}}^{\nu}\frac{1}{z}+T_{1}^{\nu}z+
T^{\nu}_{2}z^2=\mu.
\end{equation}
It is clear that the algebraic equation gives four solutions of $z$ for the given chemical potential and the transverse momentum. However, not all solutions are allowed. For the infinite lattice, the wave function must remain finite at infinities, $|\psi(\infty)|<\infty$ and $|\psi(-\infty)|<\infty$. It implies that only $|z|=1$ solutions are allowed. These are the plane-wave solutions with real momentum defined as $z= e^{ik}$. However, for an open boundary with zigzag shape, the boundary conditions change to 
\begin{equation}\label{zbc}
\psi(-1)=0,\hspace{0.3cm}\psi(0)=0,\hspace{0.3cm}
\left|\psi(\infty)\right| < \infty.
\end{equation}
Thus, $|z| \leq 1$ is required to keep the wave function finite which is less strict than the $|z|=1$ criterion for translational invariant systems. However, we have additional two boundaries conditions at $x=0, -1$, the edge state does not always exist, unless we have enough $|z_{\gamma}| \leq 1$ zero-modes to construct the edge states.

\begin{figure}
\begin{center}
\includegraphics[width=4.cm]{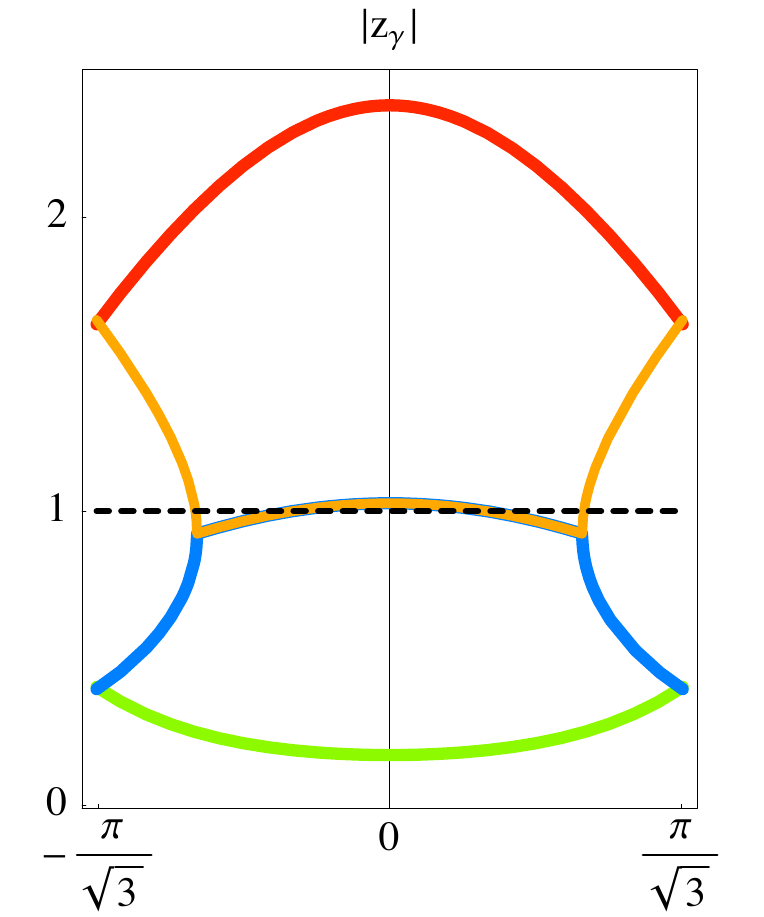}
\includegraphics[width=4.cm]{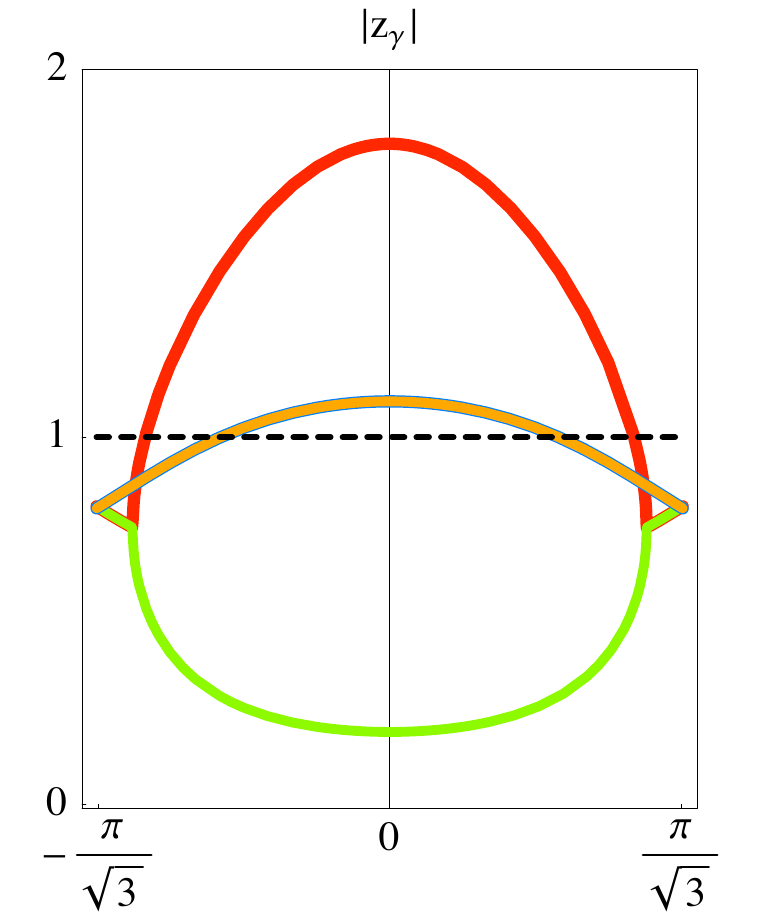}
\includegraphics[width=4.cm]{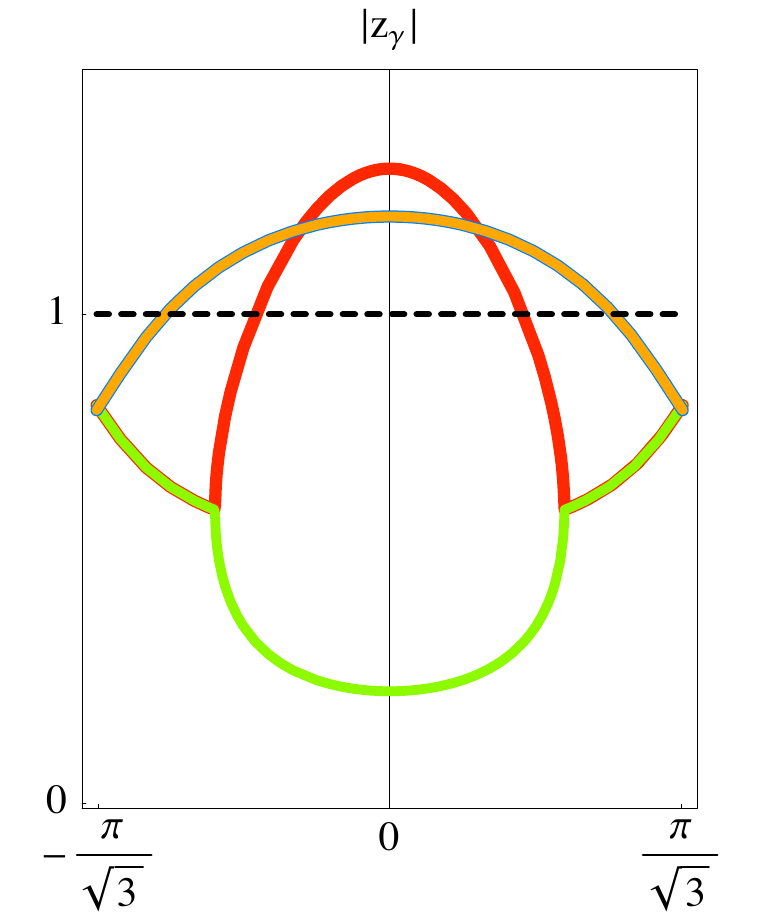}
\includegraphics[width=4.cm]{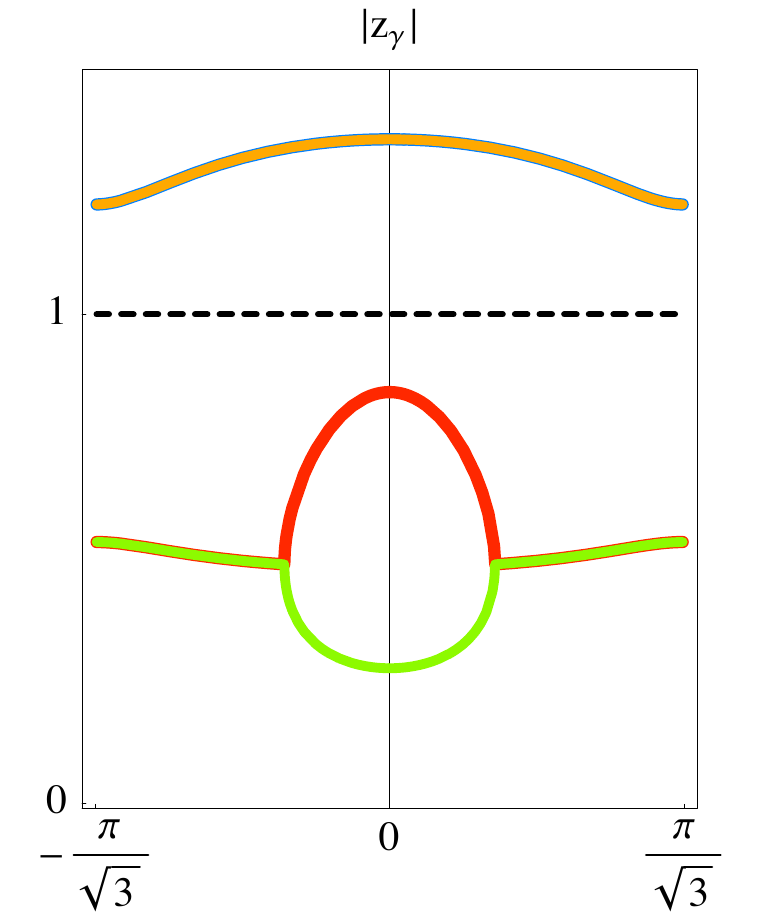}
\caption{(Color Online) The $k_y$ momentum dependence of $|z|$ for the generalized Bloch states with the $f$-wave symmetry at the zigzag edge. The lines with different colors represent four solutions of the Bloch state with the parameters, $t=1$ and $\Delta=0.4$. The chemical potentials in the top-left and top-right figures are $\mu/t=4$ and $\mu/t=1$ respectively. The bottom-left and bottome-right figures are for $\mu/t=-1$ and $\mu/t=-2.5$.}\label{zsolution}
\end{center}
\end{figure} 

Here comes the simple counting. If all of the four solutions satisfy $|z_{\gamma}| \leq 1$, we can construct two edge states. If three solutions are found, one edge state can be constructed. Otherwise, there will be no edge state. In the case of the $f$-wave pairing symmetry, we plot the magnitude of the solutions $|z_{\gamma}|$ as a function of the transverse momentum $k_y$ in Fig. \ref{zsolution}. The $z$-plot sensitively depends on the chemical potential $\mu$.

\begin{figure}
\begin{center}
\includegraphics[width=7.8cm]{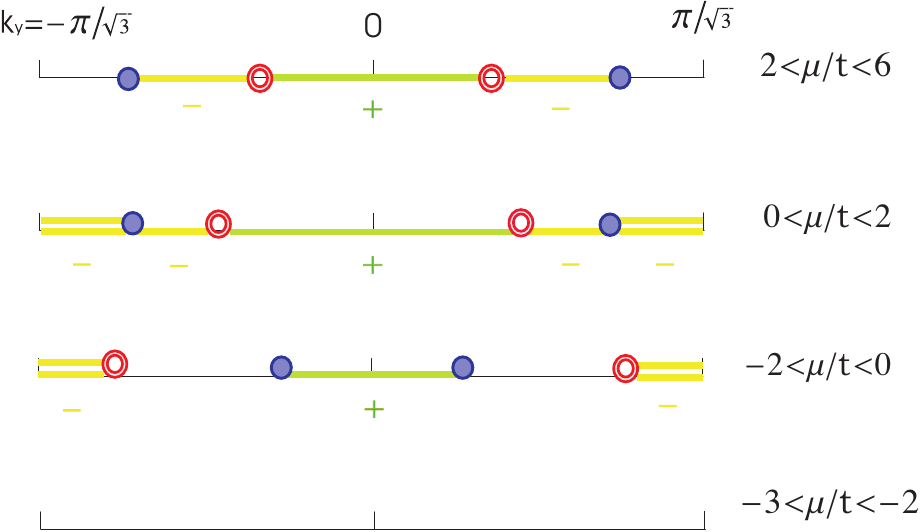}
\caption{(Color Online) Phase diagram for AES with the $f$-wave pairing at different chemical potentials. The single point circle mark the nodal points without degeneracy while the double circle denote the two-fold degenerate nodal points. The green/yellow colors denote the SUSY parity $\pm 1$ and the single/double lines mean one/two-fold degenerate edge states.
}\label{PDZf}
\end{center}
\end{figure}

Now we would like to explain how to obtain the phase diagram for AES from the $z$-plots. We start with the first $z$-plot (upper left) in Fig.~\ref{zsolution} where the chemical potential is $\mu/t=4$ and the pairing potential is $\Delta/t=0.4$. There are four intersections with $|z|=1$ dashed line. These are the nodal points. At larger momentum, the $|z|=1$ dashed line intersects with one solution (orange line) and gives rise to the nodal point. At small momentum, the dashed line intersects with two degenerate solutions (orange and blue lines) at the same time and corresponds to a pair of degenerate nodal points. These nodal points correspond to the single and double circles in the phase diagram. Now we can proceed to determine how many edge states $|\Psi_-\rangle$ with negative Witten parity can be found. Near the zone boundary $k_y=\pi/\sqrt{3}$, there are two solutions (green and blue lines) with $|z| \leq 1$. Since there are two constraints from the zigzag boundary, no edge state can be constructed. Passing the nodal point, there are three solutions (green, blue and orange lines) and thus one edge state starts to emerge. The AES with negative Witten parity is marked by yellow color in the phase diagram. Moving toward to zone center, the number of desired solution reduces to one (green line) after passing the two-fold degenerate nodal point. Thus, no edge state in presence in this regime. Due to the parity symmetry in $y$-direction, the phase diagram is symmetric when $k_y \to -k_y$.

What about the edge state $|\Psi_+ \rangle$ with positive Witten parity? One should repeat the derivation for the characteristic equation and look for $|z| \leq 1$ solutions to construct the edge states again. However, there is some symmetry hidden in the algebraic equation and the repetition is not necessary. Since the matrices $\mathcal{A}$ and $\mathcal{A}^\dag$ are hermitian conjugate to each other, the algebraic equation for the positive Wittien parity modes can be obtained by replacing $z \to 1/z$ in Eq.~\ref{zsol1}. That is to say, the decaying solutions for positive Witten parity can be calculated from the $|z| \geq 1$ solutions in Eq.~\ref{zsol1}. This relation is very helpful in constructing the remaining part of the phase diagram. Near the zone center in the first $z$-plot (upper left) in Fig.~\ref{zsolution}, there are three $|z| \geq 1$ solutions and correspond to one edge state with positive Witten parity. In other regimes, no such edge state exists. Combining the results for both Witten parities, the first part of the phase diagram in Fig.~\ref{PDZf} is obtained.

\begin{figure}
\begin{center}
\includegraphics[width=7cm]{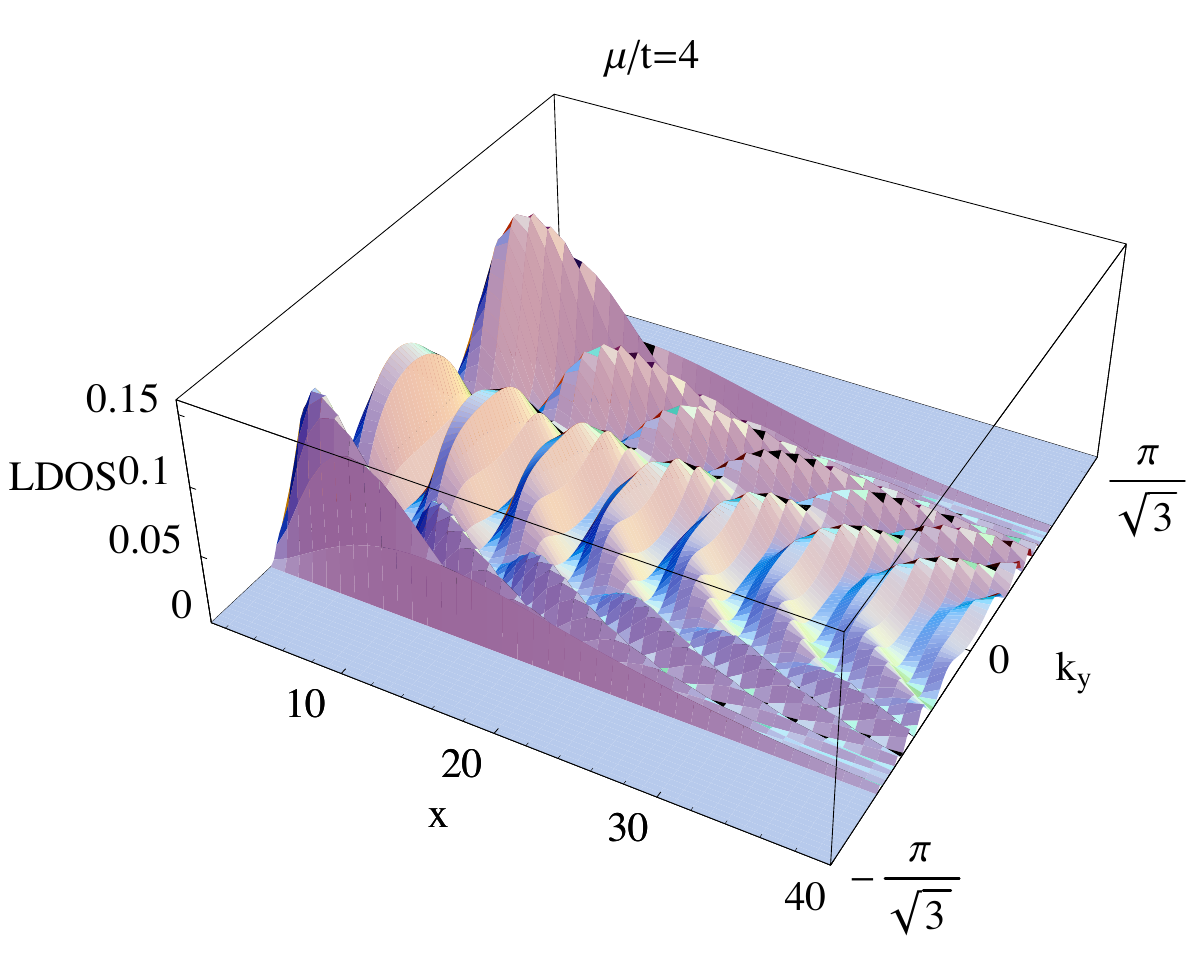}
\includegraphics[width=6.cm]{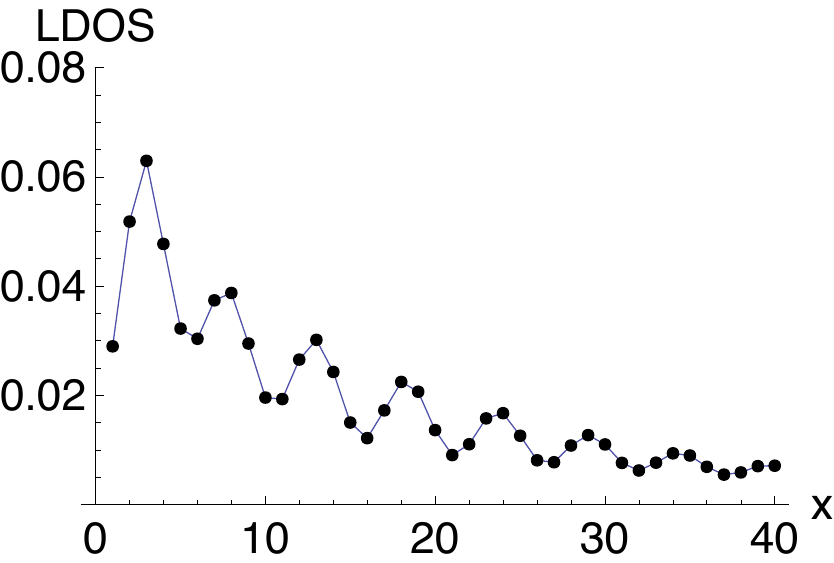}
\caption{(Color Online) LDOS for the $f$-wave pairing (top figure) along the $x$-direction at different transverse momentum $k_y$ at the chemical potential $\mu/t=4$. The bottom figure shows the spatial trend of the integrated LDOS over the Brillouin zone that can be measured directly from STM experiments.}\label{LDOS1}
\end{center}
\end{figure}

\begin{figure}
\begin{center}
\includegraphics[width=7cm]{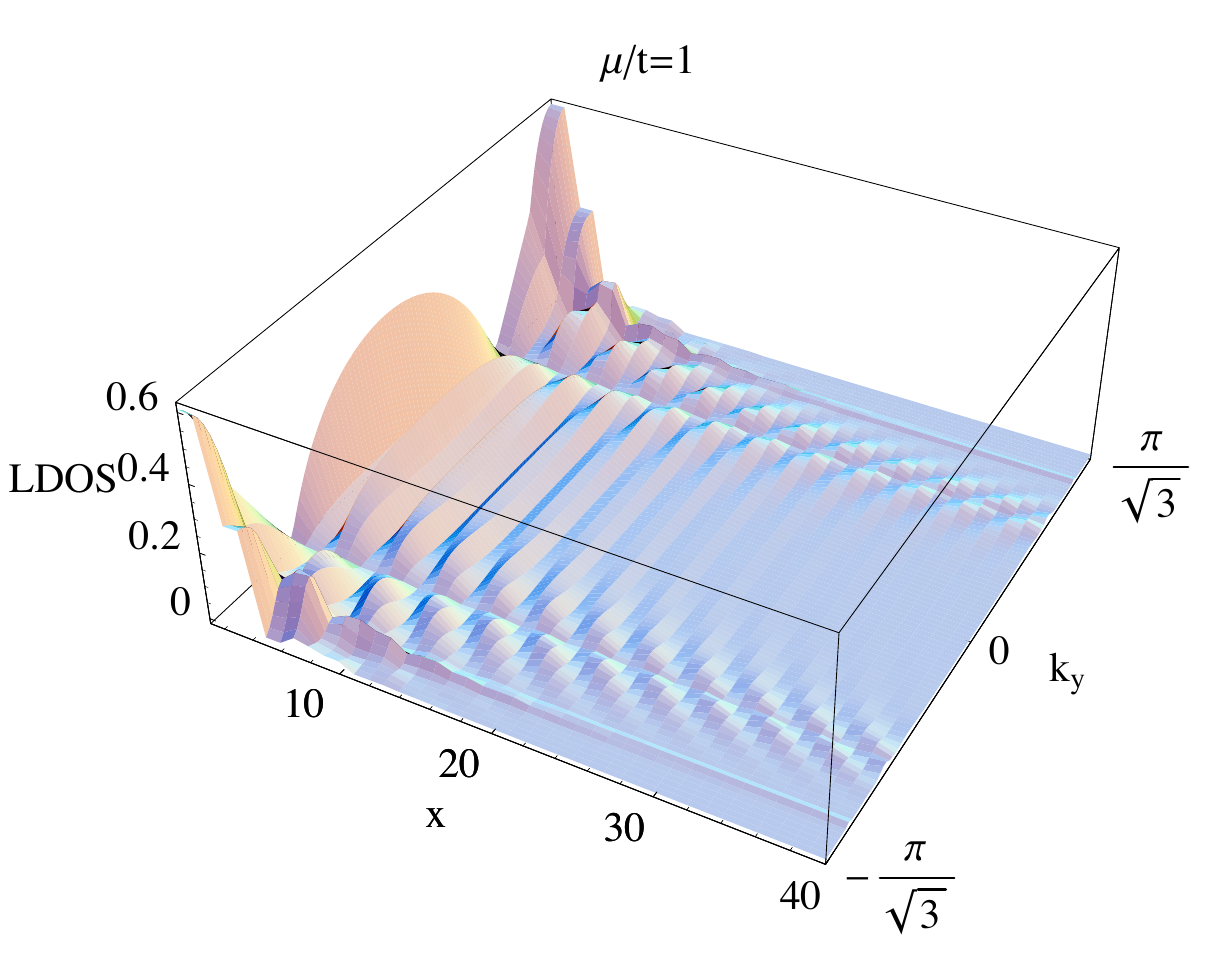}
\includegraphics[width=6.cm]{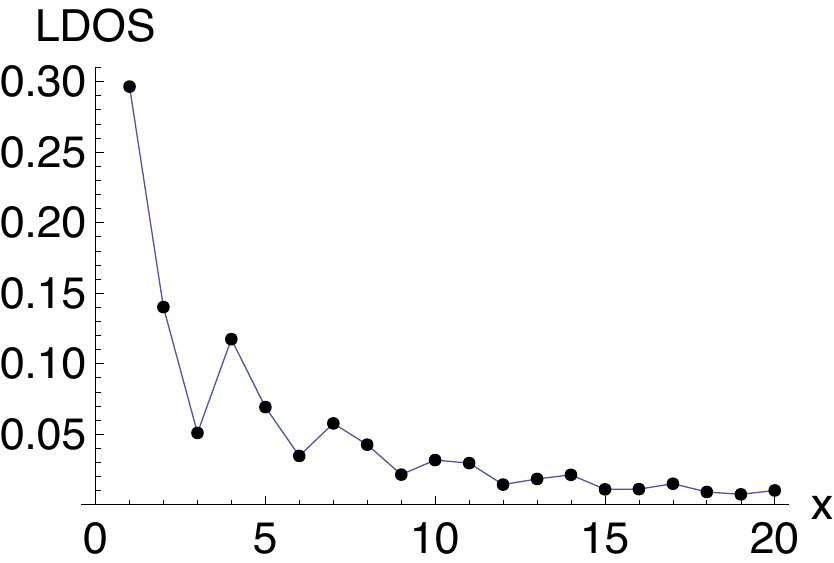}
\caption{(Color Online) momentum resolved (top) and the integrated (bottom) LDOS for the $f$-wave pairing at the chemical potential $\mu/t=1$.}\label{LDOS2}
\end{center}
\end{figure}

\begin{figure}
\begin{center}
\includegraphics[width=7.cm]{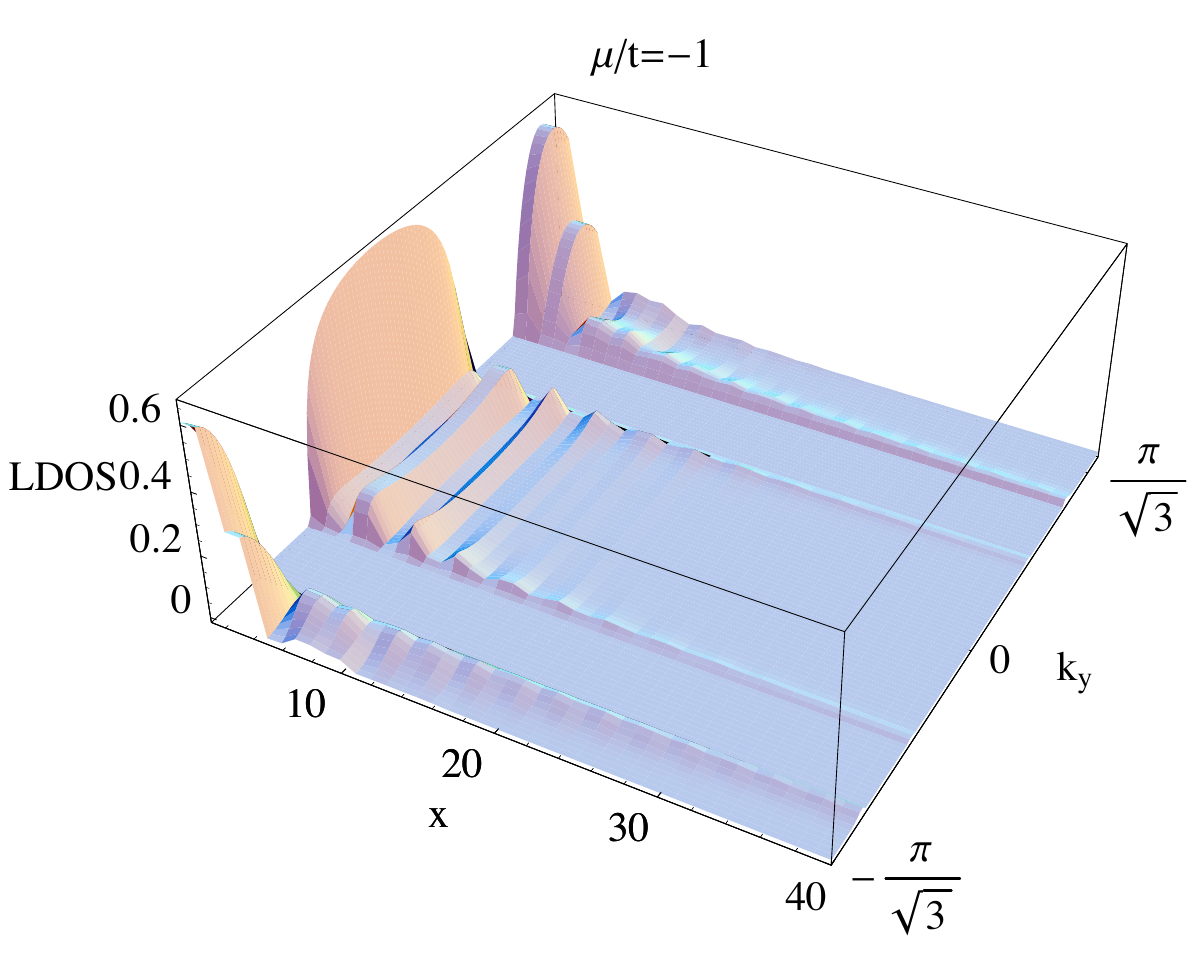}
\includegraphics[width=6.cm]{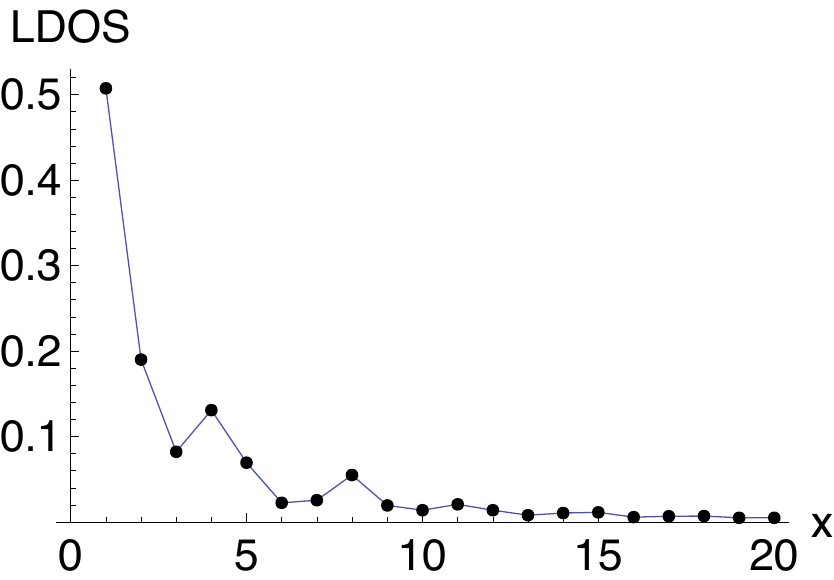}
\caption{(Color Online) momentum resolved (top) and the integrated (bottom) LDOS for the $f$-wave pairing at the chemical potential $\mu/t=-1$.}\label{LDOS3}
\end{center}
\end{figure}

Since we compute the value of $z$ for each transverse momentum $k_y$, the quasiparticle wave function of Bloch states can be obtain straightforwardly. Thus, in addition to the phase diagram, we can also compute the LDOS of the edge state at specific transverse momentum $k_y$. We can also integrate over the Brillouin zone to obtain the spatial profile for LDOS that can be measured directly in the STM experiments. Furthermore, the momentum-resolved LDOS provides additional information about the enhanced spectral weight of the quasi-particles at specific transverse momenta. Thus, we can predict the evolution of the so-called ``hot spots" in the FT-STS experiments.

Let us elaborate on the physical properties of the AES now. In order to visualize these edge states better, we calculate the local density of states $D(x,k_y)=\sum_j|\psi_j(x)|^2\delta(E)$ versus transverse momentum $k_y$, as shown in the top panels of Figs. \ref{LDOS1}, \ref{LDOS2} and \ref{LDOS3} at different chemical potentials. Noted that the edge states merge into the bulk at the nodal points and the weighting of the LDOS is suppressed to zero. Furthermore, the lattice approach reveals a much richer spatial structure in comparison with the conventional Andreev equations in the continuous limit. For instance, the LDOS has a strong dependence on the transverse momentum with transparent peak structures. For $2<\mu/t<6$, there are three peaks separated by the nodal points and the peak positions change with the chemical potential. At $\mu/t=2$, the outer peaks move to the boundary of Brillouin zone and merge into one. Therefore, for $0 < \mu/t<2$, there are only two peaks located at the center and the boundary of the Brillouin zone and the locations of the peaks do not change with the chemical potential. Further reducing the chemical potential to the regime $ -2 < \mu/t <0$, the relative weights of the peaks change but the locations remain fixed.

By integrating over the Brillouin zone, we can compute the spatial profile of the LDOS in coordinate space as shown in the bottom panels of Figs. \ref{LDOS1}, \ref{LDOS2} and \ref{LDOS3} at different chemical potentials. On top of the decaying trend, the LDOS also shows non-trivial oscillation due to quantum interferences due to different zero modes. These oscillations can only be captured faithfully in the lattice approach. For instance, at $\mu/t=4$, the LDOS at the outmost edge site is not the largest as one would naively expect so in the continuous theory. Furthermore, the decay length is smaller as the chemical potential decreases.

\begin{figure}
\begin{center}
\includegraphics[width=7.3cm]{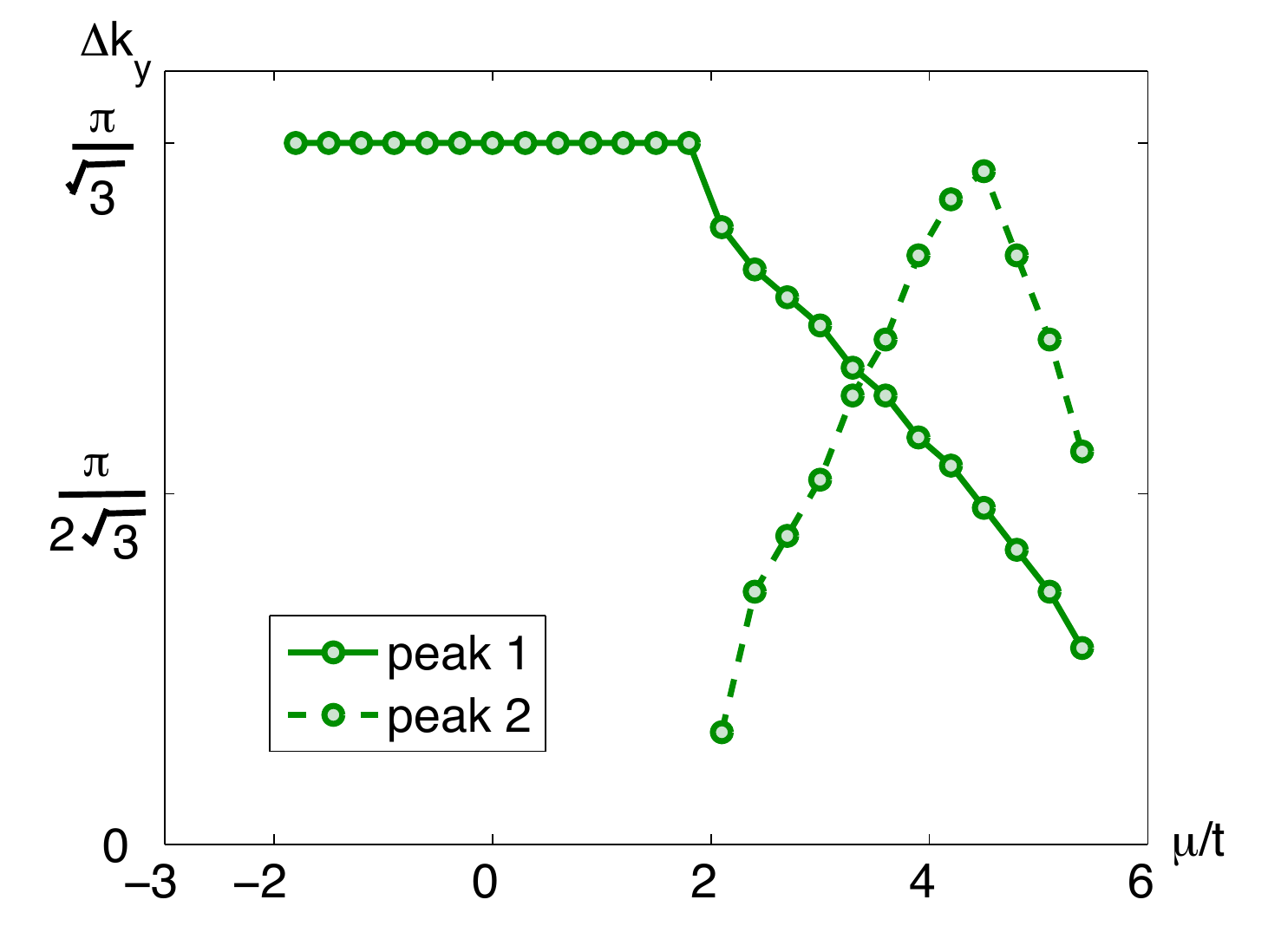}
\caption{(Color Online) Predicted hot spots in FT-STS experiment for the $f$-wave symmetry near the zigzag edge. The hot spots are obtained by locating the momentum difference between large peaks in LDOS. After $\mu/t<2$, the second peak disappears since the outer peaks merge into one at the boundary of Brillouin zone.}\label{STMzf}
\end{center}
\end{figure}

The momentum-resolved LDOS can also help us to determine the hot spots due to quasi-particle scattering/interferences in FT-STS experiments. By Fourier analysis of the STM data, the momentum transfer between quasi-particle scattering is revealed. The momentum transfer associated with the scattering process between peaks in LDOS will emerge after Fourier transformation. In Fig.~\ref{STMzf}, the momentum transfers between peaks in LDOS are plotted versus the chemical potential. For $\mu/t >2$, there are three peaks giving two specific momentum transfers. For $\mu/t<2$, there are only two peaks located at the center and the boundary of the Brillouin zone. Thus, the momentum transfer is always $\pi/\sqrt{3}$ that is half of the Brillouin zone.

\subsection{$d_{xy}$-wave pairing}

\begin{figure}
\begin{center}
\includegraphics[width=6.5cm]{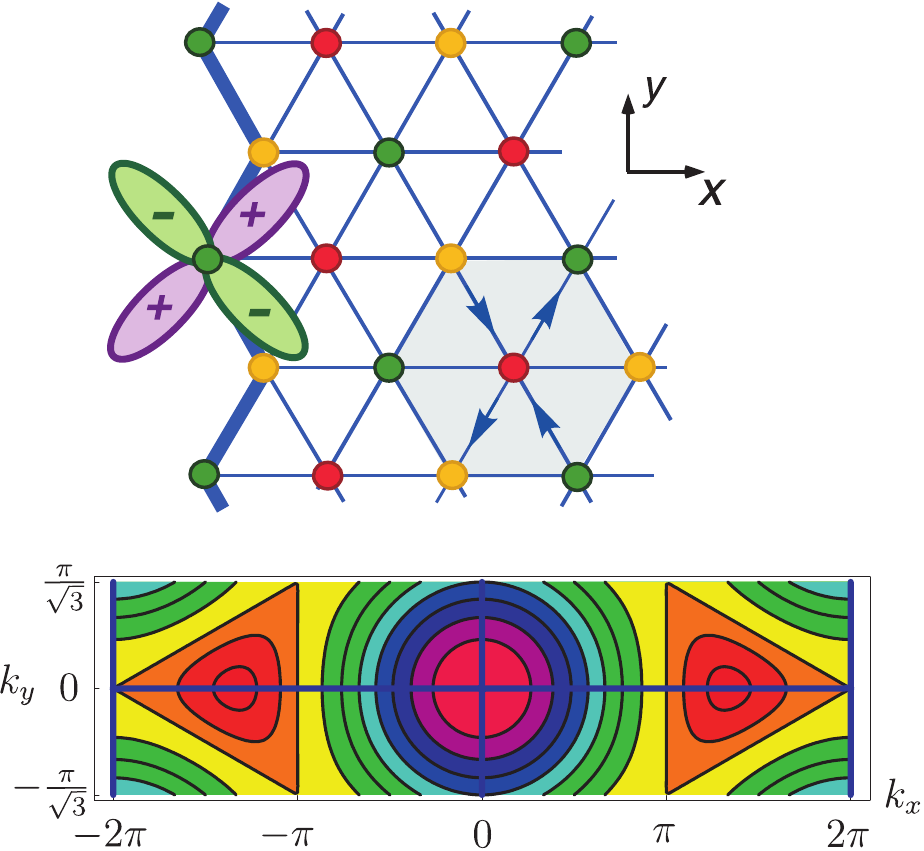}
\caption{(Color Online) Gap function with the $d_{xy}$ symmetry at the zigzag edge of a triangular lattice. The bottom figure represents the reshaped Brillouin zone, nodal lines and nodal points for the $d_{xy}$ symmetric gap function with the same convention as explained in the $f$-wave case.
}\label{PSZdxy}
\end{center}
\end{figure}

For easier experimental comparisons, we would also work out some other pairing symmetries explicitly. Since the derivations are rather similar, we would skip the repeated parts and concentrated on the different outcomes. Now we turn to the $d_{xy}$ pairing symmetry at the zigzag edge. For $d_{xy}$ pairing symmetry, Cooper pairs form spin singlets and the gap function in the coordinate space is thus symmetric, $\Delta(\textbf{r}, \textbf{r}')=\Delta( \textbf{r}', \textbf{r})$. Again, within the tight-binding approximations, the pairing potential is rather simple, $\Delta(\textbf{r}, \textbf{r}')=\Delta\sin2\theta$, with relative angle $\theta=2n\pi/6$ where $n$ is an integer. The nodal lines, satisfying the constraint $\sin(k_x/2)\sin(\sqrt{3}k_y/2)=0$, are shown in the reconstructed Brillouin zone in Fig. \ref{PSZdxy}. The Hamiltonian for the hopping is identically the same so that we do not put it down again. On the other hand, the pairing potential consists of another semi-infinite matrix $\mathcal{P}_{d}$,
\begin{eqnarray}
\mathcal{P}_{d}=\left(
\begin{array}{cccccc}
0 & -i\Delta_2 & 0 & 0 & 0 & \cdots \\
i\Delta_2& 0 & -i\Delta_2 & 0 & 0 & \cdots \\
0 & i\Delta_2 & 0 & -i\Delta_2& 0 & \cdots \\
0& 0 & i\Delta_2 & 0 & -i\Delta_2 & \cdots \\
\cdot & \cdot & \cdot & \cdot & \cdot & \cdots \\
\cdot & \cdot & \cdot & \cdot & \cdot & \cdots
\end{array}
\right),
\end{eqnarray}
with the effective 1D pairing potential $\Delta_2=\sqrt{3}\hspace{0.1cm}\Delta\sin(\sqrt{3}k_y/2)$. Note that the next nearest-neighbor pairing potentials are absent due to the nodal structure of $d_{xy}$-wave pairing symmetry along the $x$-direction (see Fig. \ref{PSZdxy}). Making use of $\mathcal{P}_{d}=\mathcal{P}_{d}^{\dag}$, a unitary transformation is devised,
\begin{eqnarray}
\Psi_{d}(x,k_{y})= \left[
\begin{array}{c}
c_{\downarrow}(x,k_{y})- i\: c_{\uparrow}^{\dag}(x,-k_{y})\\
c_{\downarrow}(x,k_{y})+ i\: c_{\uparrow}^{\dag}(x,-k_{y}) 
\end{array}\right],
\end{eqnarray}
to bring the BdG Hamiltonian into the SUSY form in Eq.(\ref{DiracZ}). Although the pairing symmetry is different, the structure of the SUSY Hamiltonian remain the same form. After some algebra, the off-diagonal components of the semi-infinite matrix $\mathcal{A}_d$ are,
\begin{eqnarray}
\nonumber T_{1,\bar{1}}^d&=&2t\cos\left(\frac{\sqrt{3}}{2}k_y\right)\mp\sqrt{3}\Delta\sin\left(\frac{\sqrt{3}}{2}k_y\right),\\
T_{2,\bar{2}}^d&=&t.
\end{eqnarray}

\begin{figure}
\begin{center}
\includegraphics[width=7.8cm]{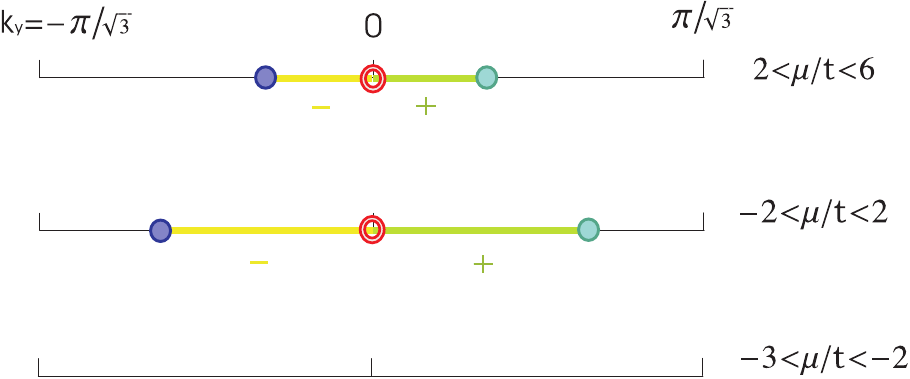}
\caption{(Color Online) Phase diagram for AES with the $d_{xy}$ pairing in the presence of the zigzag edge. The meanings of the labels are the same as in the $f$-wave case. }\label{PDZd}
\end{center}
\end{figure}

As mentioned before, the hidden SUSY in the BdG Hamiltonian makes the zero-energy modes nodal for all pairing symmetries. Repeating the same calculations, the $z$-plots are obtained at different chemical potentials. The only differences are the matrix elements $ T^d_{1,\bar{1}}$ and  $T^d_{2,\bar{2}}$ due to different pairing symmetry. By counting the decaying modes with $|z| \leq 1$, we can construct the phase diagram for AES with $d_{xy}$ pairing symmetry as shown in Fig. \ref{PDZd}. The phase diagram for $d_{xy}$ symmetry appears to be much simpler since the number of nodes are reduced and the only double node lie in $k_y=0$. Starting from the regime $2<\mu/t <6$, there exists an edge state with the positive/negative Witten parity depending on the sign of the transverse momentum. When reaching $\mu/t=2$, the nodal points move to the boundary of the reshaped Brillouin zone so that the edge state exists for every transverse momentum. Further reducing the chemical potential to the regime $-2<\mu/t<2$, the nodal points move backward to the center again. For $-3<\mu/t<-2$, no edge state can be found. It is worth mentioning that the nodal point connecting edge states with opposite Witten parities must be two-fold degenerate by simple counting. Finally, we also calculated the hot spots at different chemical potentials, as shown in Fig. \ref{STMdp}, which can be measured in FT-STS experiment.

\subsection{$p_x$-wave paring}

\begin{figure}
\begin{center}
\includegraphics[width=6.5cm]{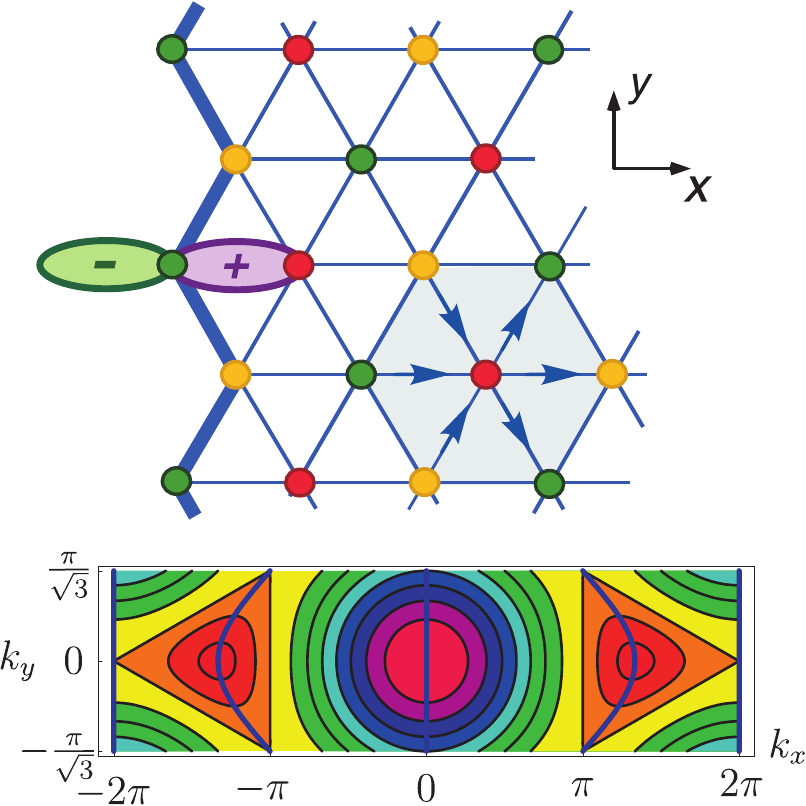}
\caption{(Color Online) Gap function with the $p_{x}$ symmetry at the zigzag edge of a triangular lattice. The bottom figure represents the reshaped Brillouin zone, nodal lines and nodal points for the $p_{x}$ symmetric gap function with the same convention as explained in the $f$-wave case.
}\label{PSZpx}
\end{center}
\end{figure}

We come to the last case at the zigzag edge -- the $p_x$ pairing symmetry. Within the tight-binding approximations, the gap function is $\Delta(\textbf{r}, \textbf{r}')=\Delta\sin\theta$, with relative angle $\theta=2n\pi/6$ where $n$ is an integer. The nodal lines in the momentum space, as shown in Fig. \ref{PSZpx}, is determined by the constraint, $\left[2\cos(k_x/2)+\cos(\sqrt{2}k_y/2)\right]\sin(k_x/2)=0$. Following the same steps, the semi-infinite matrix $\mathcal{P}_{p}$ is
\begin{eqnarray}
\mathcal{P}_{p}=\left(
\begin{array}{cccccc}
0 & \Delta_3 & \Delta & 0 & 0 & \cdots \\
-\Delta_3& 0 & \Delta_3 & \Delta & 0 & \cdots \\
-\Delta & -\Delta_3 & 0 & \Delta_3 & \Delta & \cdots \\
0& -\Delta & -\Delta_3 & 0 & \Delta_3 & \cdots \\
\cdot & \cdot & \cdot & \cdot & \cdot & \cdots \\
\cdot & \cdot & \cdot & \cdot & \cdot & \cdots
\end{array}
\right)
\end{eqnarray}
with the effective gap potential $\Delta_3=\Delta\sin\left(\sqrt{3}k_y/2\right)$. It is clear that $\mathcal{P}_{p}=-\mathcal{P}_{p}^{\dag}$. Since the semi-infinite matrix $\mathcal{P}_{p}$ share the same property as $\mathcal{P}_{f}$ for the $f$-wave pairing, the same basis, Eq.(\ref{fbasis}), can utilized to bring the BdG Hamiltonian into the SUSY form. 

\begin{figure}
\begin{center}
\includegraphics[width=7.8cm]{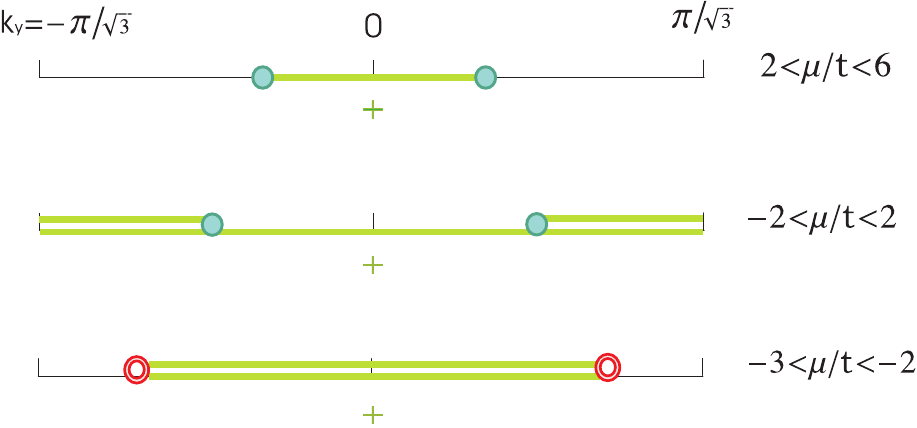}
\caption{(Color Online) Phase diagram for AES with the $p_{x}$ pairing in the presence of the zigzag edge. The meanings of the labels are the same as in the $f$-wave case.}\label{PDZp}
\end{center}
\end{figure}

\begin{figure}
\begin{center}
\includegraphics[width=7.3cm]{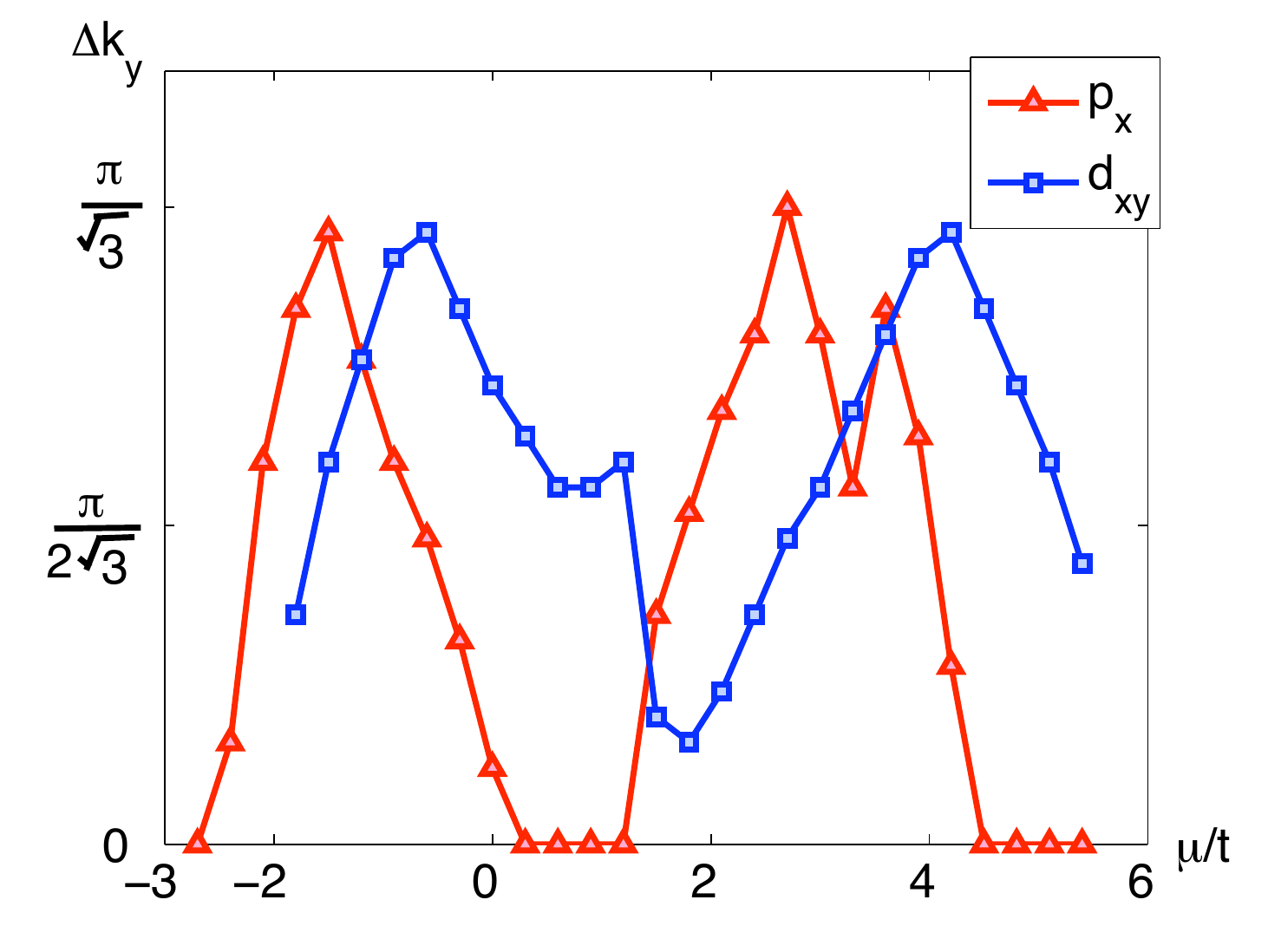}
\caption{(Color Online) Predicted hot spots in FT-STS experiment for the $d_{xy}$ or $p_x$ pairing symmetries near the zigzag edge. The hot spots are obtained by locating the momentum difference between large peaks in LDOS.
}\label{STMdp}
\end{center}
\end{figure}

After some algebra, the matrix elements of the semi-infinite matrix $\mathcal{A}_p$ in Eq.(\ref{DiracZ}) can be computed,
\begin{eqnarray}\label{MEpx}
\nonumber T_{1,\bar{1}}^p&=&(2t\pm\Delta)\cos\left(\frac{\sqrt{3}}{2}k_y\right),\\
T_{2,\bar{2}}^p&=&t\pm\Delta.
\end{eqnarray}
Following the same steps to obtain the $z$-plot, we can count the number of decaying modes with $|z| \leq 1$. The same construction leads to the phase diagram of AES for the $p_x$ pairing symmetry as plotted in Fig. \ref{PDZp}. Although we do not show the momentum-resolved LDOS for the present case, it can be computed in a similar way as for the $f$-wave pairing. Fig. \ref{STMdp} shows the evolution of the momentum transfer between the peaks in LDOS at different chemical potentials and can be compared with the hot spots in the FT-STS measurements.

As Fig. \ref{PDZp} shows, for the $p_x$ pairing symmetry, all edge states live in the null space of the semi-inifinite matrix $\mathcal{A}_{p}^{\dag}$, which are rather different from the $f$- and $d$-wave symmetries. That is to say, only AES with positive Witten parity (according to the our convention here) exists! The qualitative difference arises from the sign of the gap function across the open boundary. For the $p_x$ pairing symmetry, the pairing potentials at the edge sites all share the same sign. The pairing potential only changes signs when crossing the edge along the $y$-direction. As a result, the null space of the semi-infinite matrix $\mathcal{A}_p$ vanishes and all edge states belong to the null space of $\mathcal{A}_p^\dag$ instead. Later, we will find that it also happens for the $p_y$ pairing symmetry at the flat edge. Again, the underlying reason is that the gap function only changes signs across the open boundary of the system.    

\section{Bogoliubov-de Gennes Hamiltonian at flat edge}

\begin{figure}
\begin{center}
\includegraphics[width=8.3cm]{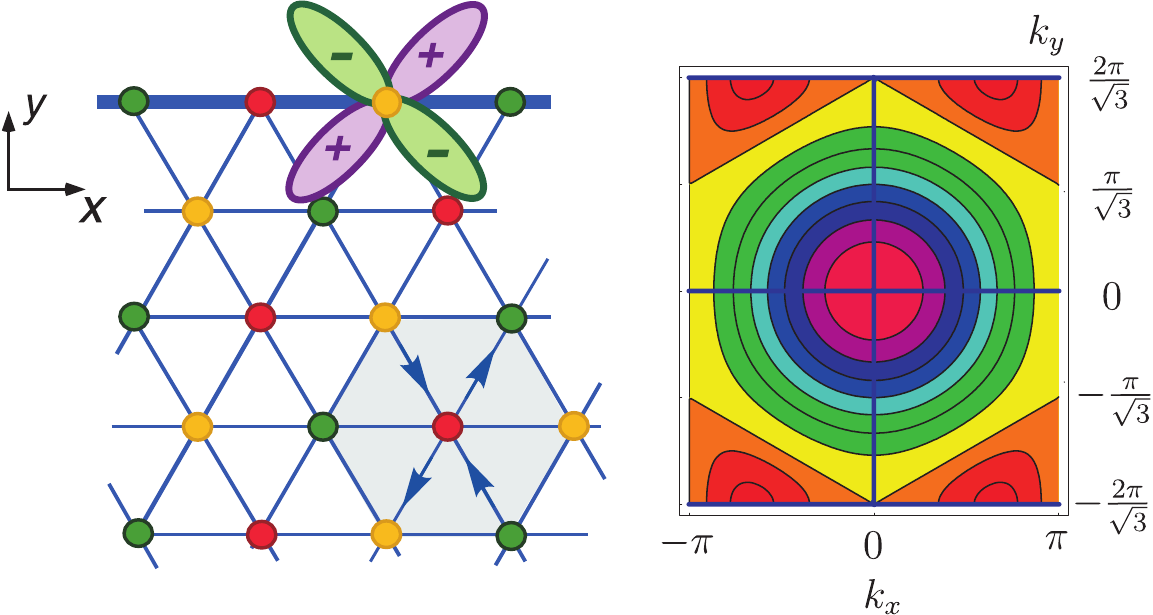}
\caption{(Color Online) Gap function with the $d_{xy}$ symmetry at the flat edge of a triangular lattice. The bottom figure represents the reshaped Brillouin zone, nodal lines and nodal points for the $d_{xy}$ symmetric gap function with the same convention as explained in the $f$-wave case.
}\label{PSFd}
\end{center}
\end{figure}

By cutting the triangular lattice in another direction (along the $x$-axis), we end up with a semi-infinite lattice with a flat edge as shown in Figs. \ref{PSFd} and \ref{PSFp}. Since the semi-infinite lattice is still translational invariant along the $x$-direction, the semi-infinte can be brought into the sum of the 1D chains by partial Fourier transformation along the edge direction. Noted that, to maintain the Fermi statistics between the lattice operators, the Brillouin zone must be reshaped in a different way as shown in Figs. \ref{PSFd} and \ref{PSFp}. In the Nambu basis, $\widetilde{\Phi}^{\dag}(k_{x},y)=\left[c^{\dag}_{\downarrow}(k_{x},y)\hspace{0.1cm},\hspace{0.1cm}c_{\uparrow}(-k_{x},y)\right]$, The BdG Hamiltonian of the $\nu$-wave pairing symmetry can be represented as,
\begin{eqnarray}\label{BdGF}
\widetilde{H}=\sum_{k_{x}}\widetilde{\Phi}^{\dag}(k_{x},y) \left(
\begin{array}{cc}
\widetilde{\mathcal{H}} & \widetilde{\mathcal{P}}_{\nu} \\
\widetilde{\mathcal{P}}^{\dag}_{\nu} & -\widetilde{\mathcal{H}}
\end{array}\right)
\widetilde{\Phi}(k_{x},y).
\end{eqnarray}
Here $\widetilde{\mathcal{H}}$ is a semi-infinite matrix for the effective hopping in the 1D chains labeled by different momentum $k_x$, 
\begin{eqnarray}
\widetilde{\mathcal{H}}=\left(
\begin{array}{cccccc}
-\widetilde{\mu} & \widetilde{t}_1& 0 & 0 & 0 & \cdots \\
\widetilde{t}_1 & -\widetilde{\mu} & \widetilde{t}_1 & 0 & 0 & \cdots \\
0 & \widetilde{t}_1 & -\widetilde{\mu} & \widetilde{t}_1 & 0 & \cdots \\
0& 0 & \widetilde{t}_1 & -\widetilde{\mu} & \widetilde{t}_1 & \cdots \\
\cdot & \cdot & \cdot & \cdot & \cdot & \cdots \\
\cdot & \cdot & \cdot & \cdot & \cdot & \cdots
\end{array}
\right),
\end{eqnarray}
with the momentum-dependent hopping amplitude $\widetilde{t}_1=2t\cos(k_x/2)$. Note that the chemical potential is renormalized, $\widetilde{\mu}=\mu-2t\cos(k_x)$ after the partial Fourier transformation. Not only the hopping matrix is different from that for the zigzag edge, the other semi-infinite matrix $\widetilde{\mathcal{P}}_{\nu}$ for the pairing potentials with the $\nu$-wave pairing symmetry would be different as well. In the following, we will study the phase diagrams of AES with different pairing symmetries near the flat edge in details.

\subsection{$d_{xy}$-wave pairing}

For the $d_{xy}$ pairing symmetry, the AES exists for both the zigzag and the flat edges. After partial Fourier transformation in the $x$-direction, the pairing potential $\widetilde{\mathcal{P}}_{d}$ in Eq.(\ref{BdGF}) can be explicitly worked out,
\begin{eqnarray}
\widetilde{\mathcal{P}}_{d}=\left(
\begin{array}{cccccc}
0 & -i\widetilde{\Delta}_1 & 0 & 0 & 0 & \cdots \\
i\widetilde{\Delta}_1 & 0 & -i\widetilde{\Delta}_1 & 0 & 0 & \cdots \\
0 & i\widetilde{\Delta}_1 & 0 &-i\widetilde{\Delta}_1 & 0 & \cdots \\
0& 0 & i\widetilde{\Delta}_1 & 0 & -i\widetilde{\Delta}_1 & \cdots \\
\cdot & \cdot & \cdot & \cdot & \cdot & \cdots \\
\cdot & \cdot & \cdot & \cdot & \cdot & \cdots
\end{array}
\right)
\end{eqnarray}
with $\widetilde{\Delta}_1=\sqrt{3}\Delta\sin(k_x/2)$. To obtain the zero-energy states, it is convenient to bring the effective Hamiltonian into the SUSY form as in the zigzag case, 
\begin{eqnarray}\label{DiracF}
\widetilde{H}=\sum_{k_{x}}\widetilde{\Psi}_{\nu}^{\dag}(k_{x},y) \left(
\begin{array}{cc}
0 & \widetilde{\mathcal{A}}_{\nu} \\
\widetilde{\mathcal{A}}_{\nu}^{\dag} & 0
\end{array}\right)
\widetilde{\Psi}_{\nu}(k_{x},y),
\end{eqnarray}
where $\nu$ denotes the pairing symmetry considered. For the flat edge, the semi-infinite matrix $\widetilde{\mathcal{A}}_{\nu}$ is simpler than that for the zigzag edge since it only has two off-diagonal rows instead of four,
\begin{eqnarray}\label{amatrixf}
\widetilde{\mathcal{A}}_{\nu}=\left(
\begin{array}{cccccc}
-\widetilde{\mu} & \widetilde{T}_{1}^{\nu} & 0 & 0 & 0 & \cdots \\
\widetilde{T}_{\bar{1}}^{\nu} & -\widetilde{\mu} & \widetilde{T}_{1}^{\nu} & 0 & 0 & \cdots \\
0 & \widetilde{T}_{\bar{1}}^{\nu} & -\widetilde{\mu} & \widetilde{T}_{1}^{\nu} & 0 & \cdots \\
0& 0& \widetilde{T}_{\bar{1}}^{\nu} & -\widetilde{\mu} & \widetilde{T}_{1}^{\nu} & \cdots \\
\cdot & \cdot & \cdot & \cdot & \cdot & \cdots \\
\cdot & \cdot & \cdot & \cdot & \cdot & \cdots
\end{array}
\right).
\end{eqnarray}
The matrix elements can be worked out explicitly from the semi-infinite matrix $\widetilde{\mathcal{P}}_{\nu}$ in Eq.~(\ref{BdGF}) which depends on the pairing symmetry. For the  $d_{xy}$ pairing symmetry, the semi-infinite matrix satisfies $\widetilde{\mathcal{P}}_{d}=\widetilde{\mathcal{P}}_{d}^{\dag}$. Thus, the unitary transformation to the SUSY form is
\begin{eqnarray}
\widetilde{\Psi}_{d}(k_{x},y)= \left[
\begin{array}{c}
c_{\downarrow}(k_{x},y)- i\:c_{\uparrow}^{\dag}(-k_{x},y)\\
c_{\downarrow}(k_{x},y)+ i\:c_{\uparrow}^{\dag}(-k_{x},y) 
\end{array}\right].
\end{eqnarray}
It is straightforward to work out the matrix elements of the semi-infinie matrix $\widetilde{\mathcal{A}}_{d}$,
\begin{eqnarray}
\widetilde{T}_{1,\bar{1}}^d&=&2t\cos\left(\frac{k_x}{2}\right)\mp\sqrt{3}\Delta\sin\left(\frac{k_x}{2}\right).
\end{eqnarray}

\begin{figure}
\begin{center}
\includegraphics[width=7.8cm]{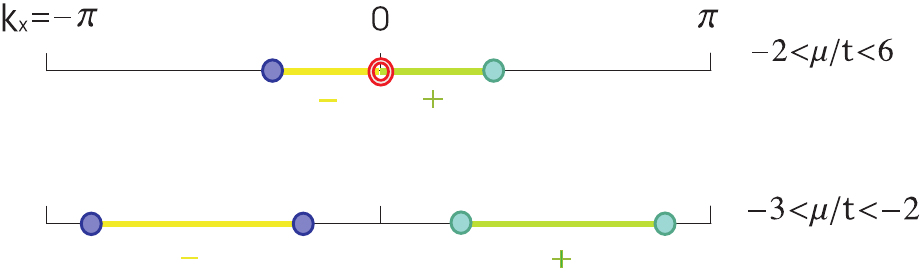}
\caption{Color Online) Phase diagram for AES with the $d_{xy}$ pairing in the presence of the flat edge. The meanings of the labels are the same as in the $f$-wave case.
}\label{PDFd}
\end{center}
\end{figure}

Again, the zero-energy modes exhibit the nodal structure and can be classified into two categories with opposite Witten parities,
\begin{eqnarray}
|\widetilde{\Psi}_{-}\rangle= \left(
\begin{array}{c}
0 \\ \widetilde{\psi}_{-}(y)
\end{array} \right),
\hspace{0.5cm} |\widetilde{\Psi}_{+}\rangle=\left(
\begin{array}{c}
\widetilde{\psi}_ {+}(y) \\ 0
\end{array} \right).
\end{eqnarray}
Here $\widetilde{\psi}_{-}(y)$ and $\widetilde{\psi}_ {+}(y)$ belong to the null space of the semi-infinite matrices $\widetilde{\mathcal{A}}_{\nu}$ and $\widetilde{\mathcal{A}}_{\nu}^{\dag}$ respectively. The edge state is constructed from the generalized Bloch theorem. For instance, the edge state with negative Witten parity is $\psi_{-}(y)=\sum_{\lambda} a_{\lambda}(z_{\lambda})^y$, where $z$ satisfies,
\begin{equation}\label{zsol2}
\widetilde{T}_{\bar{1}}^{\nu}\frac{1}{z}+\widetilde{T}_{1}^{\nu}z=\widetilde{\mu}.
\end{equation}
The above algebraic equation gives two solutions for $z$. In the presence of the flat edge, the boundary conditions are slightly different,
\begin{equation}\label{BCF}
\widetilde{\psi}(0)=0, \qquad
\left|\widetilde{\psi}(\infty) \right| < \infty.
\end{equation}
As before, only decaying modes with $|z| \leq 1$ are allowed. But, only one constraint is required at the flat edge in contrast to the two constraints for the zigzag edge. The simplification is due to the missing matrix elements $\widetilde{T}_{\bar{2}}^{\nu}$ and $\widetilde{T}_{2}^{\nu}$ at the flat edge which makes searching for the AES much easier here. The phase diagram for the AES with $d_{xy}$ pairing symmetry is shown in Fig. \ref{PDFd}. Using the Bloch wave function of those edge states, we obtain the LDOS for all transverse momenta $k_x$. Then, we can proceed to predict the sharp peaks in STM data after Fourier analysis by finding out the momentum transfer between peaks in the LDOS. The results are plotted in Fig. \ref{STMf} versus the chemical potential $\mu/t$.

\subsection{$p_y$-wave pairing}

\begin{figure}
\begin{center}
\includegraphics[width=8.cm]{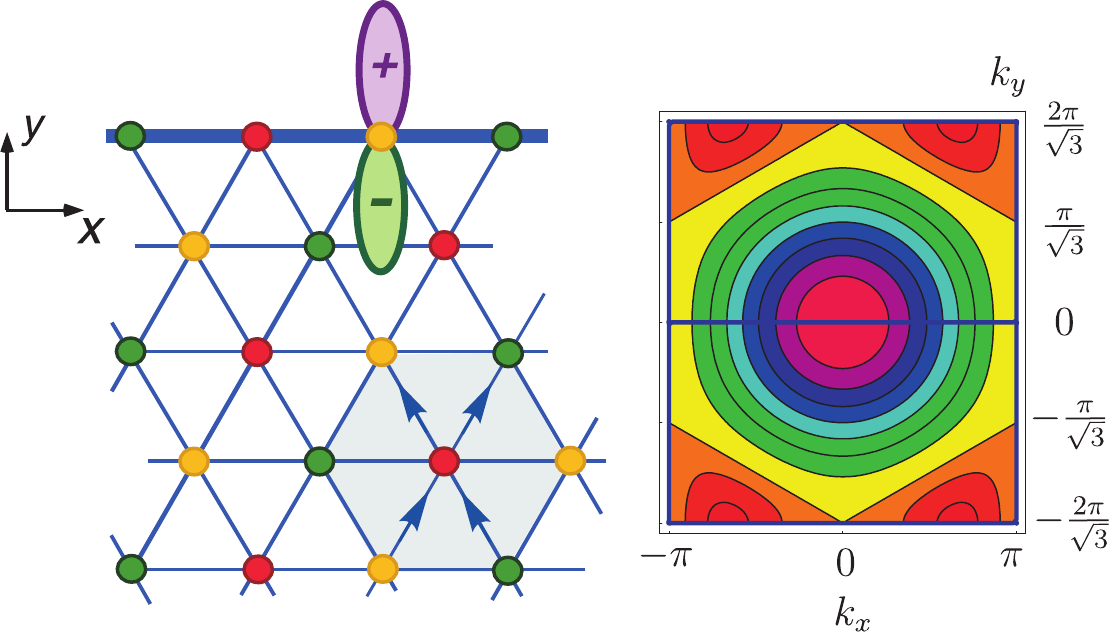}
\caption{(Color Online) Gap function with the $p_{y}$ symmetry at the flat edge of a triangular lattice. The bottom figure represents the reshaped Brillouin zone, nodal lines and nodal points for the $p_{y}$ symmetric gap function with the same convention as explained in the $f$-wave case.
}\label{PSFp}
\end{center}
\end{figure}

We now continue to study the AES with the $p_y$-wave pairing symmetry at the flat edge as shown in Fig. \ref{PSFp}. The nearest-neighbor gap amplitude of $p_y$-wave pairing takes the form, $\Delta(\textbf{r}, \textbf{r}')=\Delta\cos\theta$, with relative angle $\theta=0,\pi/3,2\pi/3,...,5\pi/3$. The nodal lines in reshaped Brilliouin zone, shown in Fig. \ref{PSFp}, are determined by the equation $\sin(\sqrt{3}k_y/2)\cos(k_x/2)=0$. Again, applying partial Fourier transformation to the gap function, we obtain the semi-infinite matrix for the pairing potential,
\begin{eqnarray}
\widetilde{\mathcal{P}}_{p}=\left(
\begin{array}{cccccc}
0 & \widetilde{\Delta}_2 & 0 & 0 & 0 & \cdots \\
-\widetilde{\Delta}_2 & 0 & \widetilde{\Delta}_2 & 0 & 0 & \cdots \\
0 & -\widetilde{\Delta}_2 & 0 & \widetilde{\Delta}_2 & 0 & \cdots \\
0& 0 & -\widetilde{\Delta}_2 & 0 & \widetilde{\Delta}_2 & \cdots \\
\cdot & \cdot & \cdot & \cdot & \cdot & \cdots \\
\cdot & \cdot & \cdot & \cdot & \cdot & \cdots
\end{array}
\right),
\end{eqnarray}
with $\widetilde{\Delta}_2=\sqrt{3}\hspace{0.1cm}\Delta\cos\left(k_x/2\right)$. Because of $\widetilde{\mathcal{P}}_{p}=\widetilde{\mathcal{P}}_{p}^{\dag}$, the Hamiltonian can be brought into SUSY form in the basis,
\begin{eqnarray}
\widetilde{\Psi}_{p}(k_{x},y)= \left[
\begin{array}{c}
c_{\downarrow}(k_{x},y)-c_{\uparrow}^{\dag}(-k_{x},y)\\
c_{\downarrow}(k_{x},y)+c_{\uparrow}^{\dag}(-k_{x},y) 
\end{array}\right].
\end{eqnarray}
After straightforward algebra, the components of the matrix $\widetilde{\mathcal{A}}_{p}$ for the $p_y$-wave pairing are
\begin{eqnarray}
\widetilde{T}_{1,\bar{1}}^p&=&(2t\pm\Delta)\cos\left(\frac{k_x}{2}\right).
\end{eqnarray}

\begin{figure}
\begin{center}
\includegraphics[width=7.8cm]{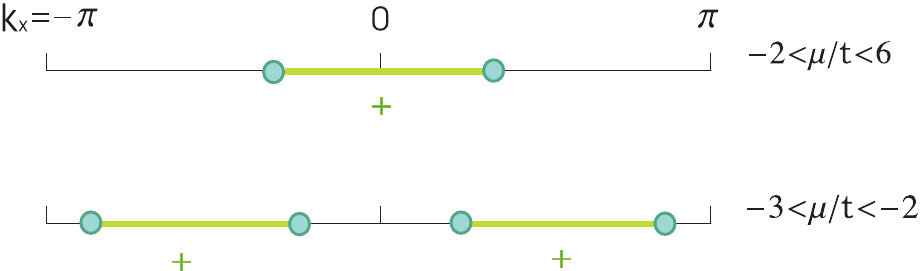}
\caption{Color Online) Phase diagram for AES with the $p_{y}$ pairing in the presence of the flat edge. The meanings of the labels are the same as in the $f$-wave case.
}\label{PDFp}
\end{center}
\end{figure}
\begin{figure}
\begin{center}
\includegraphics[width=7.3cm]{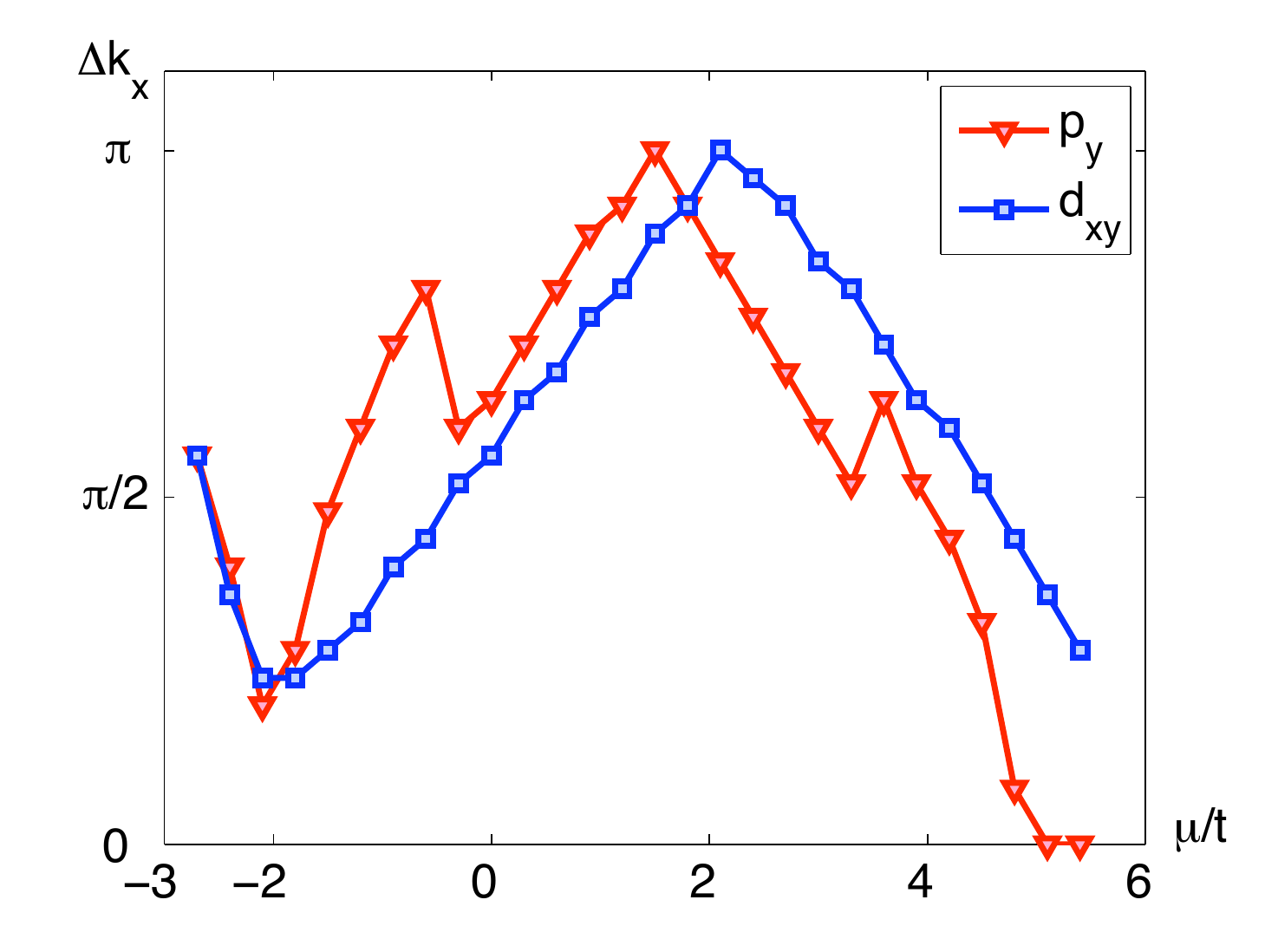}
\caption{(Color Online) Predicted hot spots in FT-STS experiment for the $d_{xy}$ or $p_y$ pairing symmetries near the flat edge. The hot spots are obtained by locating the momentum difference between large peaks in LDOS.
}\label{STMf}
\end{center}
\end{figure}

The effective 1D chains for the flat edge are universal. Thus, the whole discussions and calculations for the $d$-wave pairing with the flat edge can be applied here. Substituting the matrix elements $\widetilde{T}_{1}^p$ and $\widetilde{T}_{\bar{1}}^p$ into Eq.(\ref{zsol2}) and combining the boundary conditions of the flat edge, Eq.(\ref{BCF}), we obtain the phase diagram of AES for the $p_y$ pairing symmetry as shown in Fig. \ref{PDFp}. As we mentioned in previous section, the $p$-wave pairing potential changes sign across the edge boundary and lead to edge states with positive Witten parity only.
Finally, using the Bloch wave function of those edge states, we obtain the LDOS for all transverse momenta $k_x$. Then, we can proceed to predict the sharp peaks in STM data after Fourier analysis by finding out the momentum transfer between peaks in the LDOS. The results are plotted in Fig. \ref{STMf} versus the chemical potential $\mu/t$.

\section{discussions and conclusions}

There are simple patterns behind the phase diagrams we investigated in previous sections. For instance, the total Witten parity changes by one when crossing a single nodal point while it changes by two across the double nodal point. It seems that the global structure of the phase diagram is dictated by the nodal points. These observations are indeed correct and can be explained by the continuity of $z$-plots. However, there is something deeper about why the nodal points are so important. In the following, we would like to make use of Oshikawa's gauge argument\cite{Oshikawa00,Refael05} and explain why edge states can only start/end at the nodal points.

Suppose we wrap up the semi-infinite lattice into tubural conformation and adiabatically thread a unit flux $\Phi_0=2\pi$ through it. The flux insetion changes the Hamiltonian from $H(\Phi=0)$ to a different topological sector $H(\Phi=\Phi_0)$. If the ground state of original Hamiltonian is protected by a gap, the insertion of a unit flux also transforms the ground state from $|\Psi_0\rangle$ to $|\Psi_0'\rangle$ of the same energy. The flux insertion can be achieved by the constant vector potential $A_d=\Phi_0/L_d$ with the circumference of the tube $L_d$ in the transverse direction of $x_d$. Meanwhile, the constant vector potential commutes with the transverse momentum $\hat{P}_d$ that implies that the momentum remains constant in the whole adiabatic procedure, i.e. $\hat{P}_d|\Psi_0'\rangle=P_0|\Psi_0'\rangle$. Before being able to compare $|\Psi_0\rangle$ to $|\Psi_0'\rangle$, we need to restore the Hamiltonian to the same topological sector $H(\Phi=0)$. The required large gauge transformation is
\begin{equation}
U=\exp\left[i\frac{2\pi}{L_d}\sum_{\vec{r}}x_d\hat{n}_{\vec{r}}\right],
\end{equation}   
where $\hat{n}_{\vec{r}}$ is the electron density at $\vec{r}$. Now $UH(\Phi_0)U^{\dag}=H(0)$, so $U|\Psi_0'\rangle$ is a ground state of the original Hamiltonian $H(0)$. The momentum of the new ground state can be evaluated straightforwardly, $\hat{P}_dU|\Psi_0'\rangle=(U\hat{P}_d+[\hat{P}_d,U])|\Psi_0'\rangle=(P_0+2\pi N/L_d)U|\Psi_0'\rangle$. The total number of electrons can be separated into bulk and edge parts, $N=N_B+N_E$. The momentum shift is then, $\Delta P=2\pi N/L_d=2\pi C\nu_b+2\pi\nu_e$, with $\nu_b=N_B/V_a$ and $\nu_e=N_E/L_d$ are the filling factors of the lattice and edge respectively. The area of the system is $V_a$ and the transverse size is $C=V_a/L_d$. 

Now, let us focus on the edge part. If we fill in only one edge state with $\nu_e=1/L_d$, the momentum shift by the flux-insertion-removal trick is $\Delta P=2\pi/L_d$. The number of edge state then equal to the ground state degeneracy. Since the gauge argument holds only when the ground state is protected by a finite gap, we can move one edge state to another between the nodal points.

Another interesting perspective is to relate the existence of AES to the underlying structure of the effective 1D model.\cite{Lin05} The semi-infinte lattice can be mapped into effective 1D models. By choosing an appropriate unit cell, the 1D chain will contain only nearest-neighbor hopping described by the general Hamiltonian
\begin{eqnarray}
H={\bf C_1^{\dag}}\otimes {\bf R}+{\bf C_1}\otimes {\bf R^{\dag}}+{\bf C_0}\otimes {\bf 1},
\end{eqnarray}
where ${\bf C_1}$ is the hopping matrix connecting nearest-neighbor cells and ${\bf C_0}={\bf C_0^{\dag}}$ for the hopping within the cell. The matrices ${\bf C_0}$ and ${\bf C_1}$ are square matrices with $s$ rows, where $s$ is the number of effective lattice site in the unit cell. The semi-infinite matrix $({\bf R})_{i,i'}=\delta_{i+1,i}$ is the displacement operator on the effective 1D chain. We construct the edge states from the Bloch states, ${\bf\Phi}(i)=\sum_{\gamma}a_{\gamma}\phi_{\gamma}(z_{\gamma})^{i}$, where $z$ satisfies ${\rm det}|z{\bf C_1^{\dag}}+\frac{1}{z}{\bf C_1}+{\bf C_0}|=0$. The boundary condition is extremely simple in this representation, ${\bf C_1}{\bf\Phi}(0)=\sum_{\gamma}{\bf C_1}(a_{\gamma}\phi_{\gamma})=0$. Therefore, the number of the edge states is the dimension of the null space of of ${\bf C_1}$.

If the rank of matrix ${\bf C_1}$ is full, it means no edge state. In fact, the reflection symmetry with respect to the open boundary often implies that the rank of ${\bf C_1}$ is full. For example, the $p_y$-wave pairing symmetry at the zigzag edge, one can find out that $z$ in the Bloch state should satisfy Eq.(\ref{zsol1}) with $T_{\bar{2}}=T_{2}$ and $T_{\bar{1}}=T_{1}$. That is to say, if $z$ is a solution, $1/z$ is also a solution. Thus, except the nodal points, there are always two zero modes with $|z| <1$. Since there are also two constraints, we end up with no edge state. One can also check that the reflection symmetry makes the rank of the matrix ${\bf C_1}$ full and thus leads to no edge state.

In conclusion, we study the AES with different pairing symmetries and boundary topologies on semi-infinite triangular lattice of Na$_x$CoO$_2$$\cdot y$H$_2$O. By mapping the 2D triangular lattice to the 1D counterpart, we can obtain the phase diagram and calculate the LDOS of the AES at both zigzag and flat edges. Surprisingly, the structure of the phase diagram crucially relies on the nodal points on the Fermi surface and can be explained by an elegant gauge argument. Finally, the momentum-resolved LDOS allow us to predict the hot spots in Fourier-transformed scanning tunneling spectroscopy experiments.

We acknowledge supports from the National Science Council of Taiwan through grants NSC-96-2112-M-007-004 and NSC-97-2112-M-007-022-MY3 and also partial financial aids from the National Center for Theoretical Sciences in Taiwan.

\appendix

\section{$N=2$ Supersymmetric Quantum Mechanics}

The effective Hamiltonians in Eqs.~(\ref{DiracZ}) and (\ref{DiracF}) can be described as the $N=2$ SUSY quantum mechanics\cite{SUSY}, where $N$ is the number of supercharge operators. The two supercharge operators can be constructed explicitly
\begin{eqnarray}
Q_1=H=\left(\begin{array}{cc}0 &  \mathcal{A} \\ \mathcal{A}^{\dag} & 0\end{array}\right), \hspace{0.3cm}Q_2=\left(\begin{array}{cc}0 &  -i\mathcal{A} \\ i\mathcal{A}^{\dag} & 0\end{array}\right).
\end{eqnarray}
One can verify that all SUSY algebra is satisfied. According to the definition, the SUSY Hamiltonian is
\begin{eqnarray}
H_{SUSY}=Q_1^2=Q_2^2=\left(\begin{array}{cc} \mathcal{A}\mathcal{A}^{\dag} & 0 \\ 0 & \mathcal{A}^{\dag}\mathcal{A}\end{array}\right).
\end{eqnarray}
Once we know how to diagonalize the SUSY Hamiltonian, we can also construct the eigenstates of the supercharge operators (our goal here) as well. The SUSY algebra relates the $E>0$ (the energy of the SUSY Hamiltonian) eigenstates with the opposite {\em Witten parities} 
\begin{eqnarray}
|\Psi_{-}\rangle&=&\frac{1}{\sqrt{E}}Q^{\dag}|\Psi_{+}\rangle,\\
|\Psi_{+}\rangle&=&\frac{1}{\sqrt{E}}Q|\Psi_{-}\rangle,
\end{eqnarray}
where the complex supercharges are defined as
\begin{eqnarray}
Q&=&\frac{1}{\sqrt{2}}\left(Q_1+iQ_2\right)=\left(\begin{array}{cc}0 &  \mathcal{A} \\ 0 & 0\end{array}\right), \\Q^{\dag}&=&\frac{1}{\sqrt{2}}\left(Q_1-iQ_2\right)=\left(\begin{array}{cc}0 & 0 \\ \mathcal{A}^{\dag} & 0\end{array}\right).
\end{eqnarray}
From the transformation of the Witten parities, one can realize the energy spectrum of $Q_1$ is symmetric about $\epsilon=0$, i.e. $\epsilon=\pm|E|$. 
On the other hand, the $E=0$ states satisfy the operator equation, i.e. they live in the null space of the complex supercharge $Q$ and $Q^{\dag}$,
\begin{eqnarray}
Q|\Psi_{-}\rangle&=&0,\\
Q^{\dag}|\Psi_{+}\rangle&=&0.
\end{eqnarray}
If we do find some states satisfying the above equation, it is called good SUSY because the $E=0$ states are annihilated by supercharge. On the other hand, if we can not find any $E=0$ state. it is often referred as bad SUSY since the ground state carries non-zero supercharge\cite{SUSY}. However, for condensed matter systems, the good SUSY gives rise to the zero-energy anomaly while the bad SUSY actually makes the energy spectrum symmetric about the zero energy without anomaly.

\end{document}